\documentclass[11pt]{article}

\usepackage{epsfig,amsmath,amssymb,xspace,latexsym} 
\usepackage{graphicx}
\usepackage{wrapfig}
\usepackage{comment}
\usepackage{authblk}
\usepackage[colorinlistoftodos]{todonotes}%
\usepackage[]{hyperref} % colorlinks
\usepackage[font=small]{caption}  % modify captions
\usepackage{micro type}  % fancy spacing of fonts and end-of-lines
\usepackage[]{geometry}

\newcounter{listcounter}

\renewcommand{\thelistcounter}{\roman{listcounter}}
\newcommand{\descr}{\begin{list}{(\thelistcounter)}
{\usecounter{listcounter}
\setlength{\rightmargin}{0mm}}}
\newtheorem{lemma}{Lemma}[section]
\newtheorem{theorem}[lemma]{Theorem}

\newtheorem{definition}[lemma]{Definition}

\newtheorem{algorithm}[lemma]{Algorithm}

\newif\ifpfsymb

\newenvironment{Pf}{\par\addvspace{6pt}\addtocounter{proof}{1}
       \noindent{\bf Proof:}}{{\ifnum\value{proof}>0
                \hfill\hbox{\mbox{}\hfill$\Box$}
		\addtocounter{proof}{-1}\par\addvspace{10pt}\else\par\fi
        }}

\makeatletter
\newcounter{algo}
\def\thealgo{\@arabic\c@algo}
\def\fps@algo{tbp}
\def\ftype@algo{1}
\def\ext@algo{loa}
\def\fnum@algo{Algorithm \thealgo}
\def\algo{\@float{algo}}
\def\endalgo{\end@float}
\makeatother

\makeatletter
\def\remlab#1{\@bsphack\if@filesw {\let\thepage\relax
   \def\protect{\noexpand\noexpand\noexpand}%
\xdef\@gtempa{\write\@auxout{\string
	   \newlabel{rem:#1}{{\thelemma}{\thepage}}}}}\@gtempa
            \if@nobreak \ifvmode\nobreak\fi\fi\fi\@esphack}
\def\deflab#1{\write\@auxout{\string
	\newlabel{def:#1}{{\thelemma}{\thepage}}}}
\makeatother

{\catcode`\@=11
\gdef\setft#1#2#3{%
\def\@oddfoot{%
{\setbox0=\hbox{#1}%
\setbox1=\hbox{#3}%
\ifdim\wd0>\wd1%
\dimen0=\wd0%
\box0\hfil#2\hfil\hbox to\dimen0{\hfil\hfil\box1}%
\else \dimen0=\wd1%
\hbox to\dimen0{\box0\hfil}\hfil#2\hfil\box1\fi%
}}}}

{\catcode`\@=11
\gdef\sethd#1#2#3{%
\def\@oddhead{\vbox{\hbox to\hsize{{#1}\hfil{#2}\hfil{#3}}%
\vspace{0.06in}%
\hbox to \hsize{\hrulefill}\vspace*{-0.09in}}}
\def\@evenhead{\@oddhead}
	}
}

\renewenvironment{thebibliography}[1]
	{\begin{list}{\arabic{enumi}.}
	{\usecounter{enumi}\setlength{\parsep}{0pt}
	 \setlength{\itemsep}{0pt} 
         \settowidth
	{\labelwidth}{#1.}\sloppy}}{\end{list}}

\newcounter{arabiclistc}

\def\complaint#1{}
\def\withcomplaints{
\newcounter{mycomplaints}
\def\complaint##1{\refstepcounter{mycomplaints}%
\ifhmode%
\unskip%
{\dimen1=\baselineskip \divide\dimen1 by 2 %
\raise\dimen1\llap{\tiny -\themycomplaints-}}\fi%
\marginpar{\tiny [\themycomplaints]: ##1}}%
}

\newcounter{printertype}
\def\figprint#1{
        \ifcase \theprintertype

		\begin{center}
                 \input{#1}
		\end{center}
              \or
                 \centerline{\psfig{figure=#1.ps}}
              \else
                 \vspace*{1in}
        \fi}

\makeatletter
\long\def\@myfootnotetext#1{\insert\footins{\footnotesize
    \interlinepenalty\interfootnotelinepenalty 
    \splittopskip\footnotesep
    \splitmaxdepth \dp\strutbox \floatingpenalty \@MM
    \hsize\columnwidth \@parboxrestore
   \edef\@currentlabel{\csname p@footnote\endcsname\@thefnmark}\@makemyfntext
    {\rule{\z@}{\footnotesep}\ignorespaces
      #1\strut}}}

\def\myfootnotetext{\@ifnextchar
[{\@xfootnotenext}{\xdef\@thefnmark{\thempfn}\@myfootnotetext}}

\long\def\@makemyfntext#1{\parindent 5mm #1}

\newcounter{proof}
\def\@@meqncr{\let\@tempa\relax
    \ifcase\@eqcnt \def\@tempa{& & &}\or \def\@tempa{& &}
      \else \def\@tempa{&}\fi
     \@tempa $\Box$\addtocounter{proof}{-1}
     \global\@eqnswtrue\global\@eqcnt\z@\cr}
\@namedef{meqnarray*}{\def\@eqncr{\nonumber\@seqncr}\eqnarray}
\@namedef{endmeqnarray*}{\global\@eqcnt\tw@\@@meqncr\egroup
      \global\advance\c@equation\m@ne$$\global\@ignoretrue}

\@namedef{meqnarray}{\eqnarray}
\@namedef{endmeqnarray}{\@@meqncr\egroup
        \global\advance\c@equation\m@ne$$\global\@ignoretrue}
\def\mequation{$$\global\@ignoretrue}

\@namedef{Pf*}{\Pf}
\@namedef{endPf*}{\par\addvspace{10pt}}

\makeatother

\newif\ifdraft
\drafttrue
\draftfalse  %% comment this line to see the draft-only version
\newif\iffinal
\ifdraft \finalfalse \else \finaltrue \fi

\ifdraft
 \newcommand{\holin}[1]{{\color{red} #1 }}   
 \newcommand{\nadine}[1]{{\color{blue} #1 }} 
 \newcommand{\erik}[1]{{\color{red} #1 }} 
 \newcommand{\peng}[1]{{\color{red} #1 }} 
 \newcommand{\dw}[1]{{\color{red}#1}} 
 \newcommand{\damien}[1]{{\color{red}#1}} 
\else
 \newcommand{\holin}[1]{}    
 \newcommand{\nadine}[1]{} 
 \newcommand{\erik}[1]{} 
 \newcommand{\peng}[1]{} 
 \newcommand{\dw}[1]{} 
 \newcommand{\damien}[1]{} 
\fi

\newcommand{\config}{C}
\newcommand{\nut}{monomer}
\newcommand{\nuts}{monomers}

\newcommand{\ra}{\rightarrow}

\begin{document}

\title{Active Self-Assembly of Algorithmic Shapes and Patterns \\ in Polylogarithmic Time\thanks{Supported by NSF grants CCF-1219274, CCF-1162589, and 0832824---the Molecular Programming Project, an NSF Graduate Fellowship, and The Caltech Center for Biological Circuit Design.}}  

\author{}
\date{}
\maketitle

\begin{center} 
\vspace{-40pt}
Damien Woods${}^{1,2}$
~  
Ho-Lin Chen${}^{1,2,7}$
~ 
Scott Goodfriend${}^{3}$ \\
~ 
Nadine Dabby${}^{4}$
~
Erik Winfree${}^{1,4,5,6}$ 
~ 
Peng Yin${}^{1,5,6,8}$ \\
\vspace{10pt}
{%\footnotesize
Computer Science${}^{1}$, 
Center for Mathematics of Information${}^{2}$, \\
Department of Chemical Engineering${}^{3}$, 
Department of Computation  and Neural Systems${}^{4}$,  \\
Department of Bioengineering${}^{5}$, 
Center for Biological Circuit Design${}^{6}$,
 \\
Caltech, Pasadena, CA 91125, U.S.A.  \\
Present address: Department of Electrical Engineering, National Taiwan University${}^{7}$\\
Present address: Wyss Institute for Biologically Inspired Engineering, 3 Blackfan Circle, Boston, MA 02445${}^{8}$\\
%\\
%Email:  {\tt {\{}py, ndabby, niles, winfree{\}}@caltech.edu}. } 
}
\end{center}
\vspace{3ex}

%=======================================================================
\begin{abstract}
We describe a computational model for studying the complexity of self-assembled structures with active molecular components. Our model captures notions of growth and movement ubiquitous in biological systems. The model is inspired by biology's fantastic ability to assemble biomolecules that form systems with complicated structure and dynamics, from molecular motors that walk on rigid tracks and proteins that dynamically alter the structure of the cell during mitosis, to embryonic development where large-scale complicated organisms efficiently grow from a single cell. Using this active self-assembly model, we show how to efficiently self-assemble shapes and patterns from simple monomers. For example, we show how to grow a line of monomers in time and number of monomer states that is merely logarithmic in the length of the line. 

Our main results show how to grow arbitrary connected two-dimensional geometric shapes and patterns in expected time that is polylogarithmic in the size of the shape, plus roughly the time required to run a Turing machine deciding whether or not a given pixel is in the shape. We do this while keeping the number of monomer types logarithmic in shape size, plus those monomers required by the Kolmogorov complexity of the shape or pattern. This work thus highlights the efficiency advantages of active self-assembly over passive self-assembly and motivates experimental effort to construct general-purpose active molecular self-assembly systems.
\end{abstract}

\section{Introduction}\label{sect:intro}
We propose a model of computation that takes its main inspiration from biology. 
Embryonic development showcases the capability of molecules to compute efficiently.  
A human zygote cell contains within it a compact program that encodes the geometric structure of an organism with roughly $10^{14}$ cells, that have a wide variety of forms and roles,  with each such cell containing up to tens of thousands of proteins and other molecules with their own intricate  arrangement and functions. 
As shown in Figure~\ref{fig:embryonicgrowthrates}(a), early stages of  embryonic development demonstrate exponential~\cite{young99lue} growth rates in mass and number of cells over time, showing remarkable time efficiency.  
Not only this, but the developmental path from a single cell to a functioning organism is an  intricately choreographed, and incredibly robust, temporal and spatial manufacturing process that operates at the molecular scale.   Development is probably the most striking example of the ubiquitous process of molecular self-assembly, where relatively simple components (such as nucleic and amino acids, lipids, carbohydrates) organize themselves into larger systems with extremely complicated structure and dynamics (cells, organs, humans).  To capture the potential for exponential growth during development, an abstract model must allow for growth by insertion and rearrangement of elements as well as for  local state changes that store information that guides the process, as shown for example in Figure~\ref{fig:embryonicgrowthrates}(b).

\begin{figure}
\centering
\includegraphics[width = 0.49\textwidth]{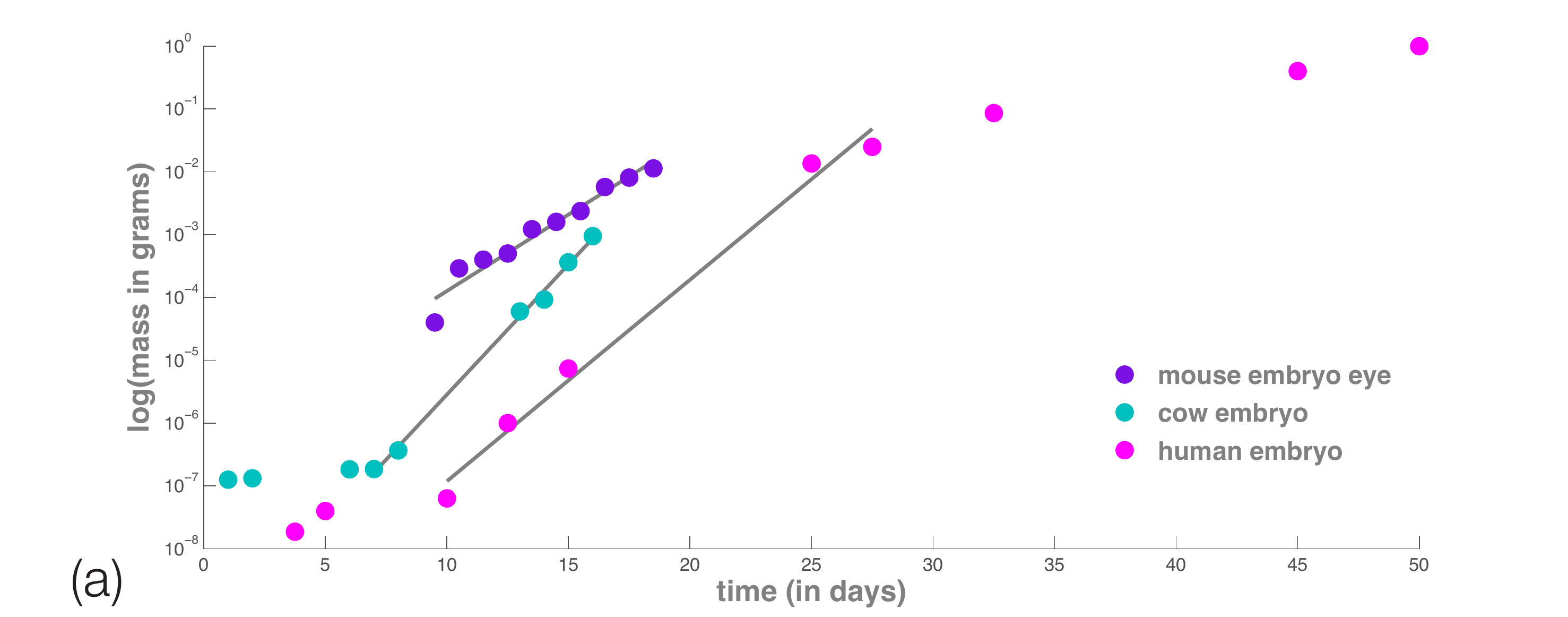} \includegraphics[width = 0.49\textwidth]{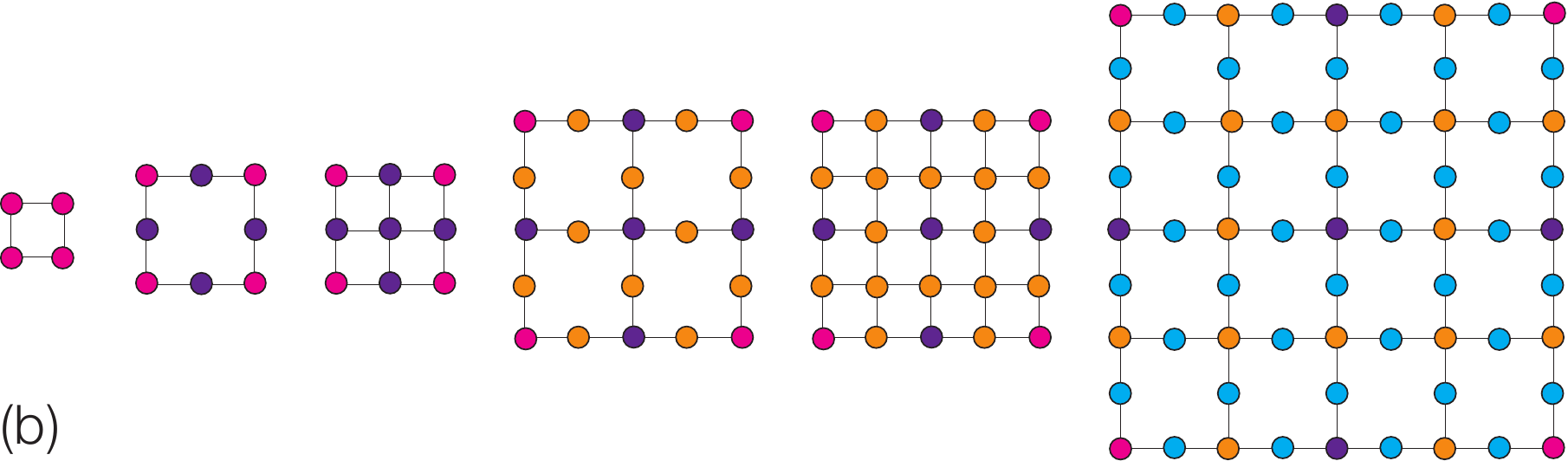}
\caption{(a) Increase in mass in embryonic mouse \cite{foster2003vivo}, cow \cite{morris2001cattle}, and human \cite{young99lue}. Gray lines highlight a time interval of exponential growth in each species. (b) An abstract example of exponential growth showing insertion and state change.\damien{perhaps we should plot number of cells as we can (1) more easily justify the exp growth rate for that, and (2) we basically do this kind of growth in nubots? I'm not sure that we know enough to say that growth in mass is best modeled exponentially. Refs are incomplete. Look at these refs again.}} 
\end{figure}\label{fig:embryonicgrowthrates}

Molecular programming, where nanoscale engineering is thought of as a programming task, provides our second motivation. The field has progressed to the stage where we can design and synthesize a range of programable self-assembling molecular systems, all with relatively simple laboratory techniques.  
For example, short DNA strands that form `tiles' can be self-assembled into larger DNA tile crystals~\cite{winfree98} that are algorithmically-patterned as counters or Sierpinski triangles~\cite{winfree04rot-sier,Winfree05bar,fujibayashi2007toward,winfree09bar}. 
This form of {\em passive self-assembly} is theoretically capable of growing arbitrarily complex algorithmically described shapes and patterns.   
Indeed, the theory of algorithmic self-assembly serves as a  guide to tell us what kinds of structures are possible, and interesting, to implement with DNA tiles.   
DNA origami can be used to create uniquely addressable shapes and patterns upon which objects can be localized within six nanometer resolution~\cite{rothemund2006folding}.   
These DNA tile and origami systems are static, in the sense that after formation their structure is essentially fixed. 
However, DNA nanotechnology has seen increased interest in the fabrication of {\em active} nanostructures that have the ability to dynamically change their structure~\cite{turberfield07bat}. Examples  include DNA-based walkers~\cite{turberfield05bat, turberfield08gre,
gu2010proximity, lund_etal2010robots,    omabegho2009bipedal,  sherman04, shin2004,   mao05tia,  pierce08yin, yin04walker},  DNA origami that  reconfigure~\cite{andersen2009self,han2010folding, marini2011revertible}, and simple structures called molecular motors that  transition between a small number of  discrete states~\cite{chakraborty2008dna,ding2006operation,Feng_Park_Reif_Yan2003,goodman2008reconfigurable,liedl2005switching,lubrich2008contractile,Mao99, Turberfield2003, pierce07ven,yurke00,zhang2011dna}.    
In these systems the interplay between structure and dynamics  leads to behaviors and capabilities that are not seen in static structures, nor in other well-mixed chemical reaction network type  systems.

Here we suggest a model to motivate engineering of molecular structures that  have complicated active dynamics of the kind seen in living biomolecular systems.  Our model combines features seen in passive DNA-based tile self-assembly, molecular motors and other active systems, molecular circuits that evolve according to well-mixed chemical kinetics, and even reaction-diffusion systems. 
The model is designed to capture the interplay between molecular structure and dynamics.  
In our model, simple molecular components form assemblies that can grow and shrink, and 
individual components  undergo state changes and move relative to each other.

The model consists of a two-dimensional grid of monomers.  
A specified set of rules, or a program, directs adjacent monomers to interact in a pairwise fashion. Monomers have internal states, and a pair of adjacent monomers can change their state with a single rule application. 
Monomers can appear and disappear from the grid. 
So far, the model can be thought of as a cellular automaton of a certain kind (that is, a cellular automaton where rules are applied asynchronously and are nondeterministic, and there is a notion of a growth front). An additional rule type allows monomers to move relative to each other. 
The movement rule is locally applied but propagates movement throughout the system in very non-local fashion. This geometric and mechanical feature distinguishes our model,  the {\em nubot} model, from previous molecular models and cellular automata, and, as we show in this paper, crucially underlies its construction efficiency. 
The system evolves as a continuous time Markov process, with rules being applied to the grid asynchronously and in parallel using standard chemical kinetics, modeling the distributed nature of molecular reactions.

The model can carry out local state changes on a grid, so it can easily simulate Turing machines, walkers and  cellular automata-like systems. We show examples of other simple programs such as robotic molecular arms that can move distance $n$ in expected time $O(\log n)$, something that can not be done by cellular automata. By using a combination of monomer movement, appearance, and state change we show how to build a line of monomers in time that is merely logarithmic in the length of the line, something that is provably
impossible in the (passive) abstract tile assembly model~\cite{AdChGoHu01}. 
We go on to efficiently build a binary counter (a program that counts builds an $n \times \log_2 n$ rectangle while counting from 0 to~$n-1$), within time and number of monomer states that are both logarithmic in $n$, where $n$ is a power of 2.  We build on these results to show that the model is capable of building wide classes of shapes exponentially faster than passive self-assembly. We show how to build a computable shape of size $\leq n \times n$ in time polylogarithmic in $n$, plus roughly the time needed to simulate a Turing machine that computes whether or not a given pixel is in the final shape. Our constructions are not only time efficient, but efficient in terms of their program-size: requiring at most  polylogarithmic  monomer types in terms of shape size, plus that required by the Kolmogorov complexity of the shape. 

For shapes of size $\leq n \times n$ that require a lot (i.e. $> n$) of time and space to compute their pixels on a Turing machine, the previous computable shape construction  requires (temporary) growth well beyond the shape boundary.  
One can ask if there are interesting structures that we can build efficiently, but yet keep all growth ``in-place'', that is, confined within the boundary. 
It turns out that colored patterns, where the color of each pixel in the pattern is computable by a polynomial time Turing machine, can be computed extremely efficiently in this way.  
More precisely, we show that $n \times n$ colored patterns are computable in expected time $O(\log^{\ell+1} n)$ and using $O(s +\log n)$ states,  where  each pixel's color is computable by a program-size $s$ Turing machine that runs in polynomial time in the length of the binary description of pixel indices (specifically, in time $O(\log^{\ell} n)$ where $\ell$ is $O(1)$).  Thus the entire pattern completes with only a linear  factor slowdown in Turing machine time (i.e.\ $\log n$ factor slowdown).  Furthermore this entire construction is initiated by a single monomer and is carried out using only local information in an entirely distributed and asynchronous fashion. 

Essentially, our constructions serve to show that shapes and patterns that are exponentially larger than their description length can be fabricated in polynomial time, with a linear number of states, in their description length, besides the states that are required by the Kolmogorov complexity of the shape.

Our active self-assembly model is intentionally rather abstract, however our results show that it captures some of the features seen in the biological development of organisms (exponential growth---with and without fast communication over long distances, active yet simple components) as well those seen in many of the active molecular systems that are currently under development (for example, DNA walkers, motors and a variety of active systems that exploit DNA strand displacement). Also, the proof techniques we use, at a very abstract level, are  informed by natural processes. In the creation of a line, a simple analog of cell division is used; division is also used in the construction of a binary counter along with a copying (with minor modifications) process to create new counter rows; in the assembly of arbitrary shapes we use analogs of protein folding and scaffold-based tissue engineering; for the assembly of arbitrary patterns we were inspired by biological development where growth and patterning takes places in a bounded region (e.g. an egg or womb) and where many parts of the development of a single embryo occur in an independent and seemingly asynchronous fashion.

Section~\ref{sect:model} contains the detailed definition of our nubot model, which is followed by a number of simple and intuitive examples in Section~\ref{sect:examples}. Section~\ref{sect:simulation} gives a polynomial time algorithm for simulating nubots rule applications, showing that the non-local movement rule is in a certain sense tractable to simulate and providing a basis for a software simulator we have developed. In Section~\ref{sect:linesSquares} we show how to grow lines and squares in time logarithmic in their size. This section includes a number of useful programming techniques and analysis tools for nubots. In Sections~\ref{sect:complexityOne} and~\ref{sect:pattern} we give our main results: building arbitrary shapes and patterns, respectively, in polylogarithmic time in shape/pattern size (plus the worst-case time for a Turing machine to compute a single pixel) and using only a logarithmic number of states (plus the Kolmogorov complexity of the shape/pattern). Section~\ref{sect:discussion} contains a number of directions for future work.

\subsection{Related work}
Although our model takes inspiration from natural and artificial active molecular processes,  it posesses a number of features seen in a variety of existing theoretical models of computation. 
We discuss some such related models here. 

The tile assembly model~\cite{winfree96win,winfree98win-thesis,winfree00rot}  formalizes the self-assembly of molecular crystals~\cite{winfree04rot-sier} from square units called tiles. This model takes the Wang tiling model~\cite{wang61wan}, which is a model of plane-tilings with square tiles, and restricts it by including a natural mechanism for the growth of a tiling from a single seed tile. Self-assembly starts from a seed tile that is contained in a soup of other tiles.  Tiles with matching colors on their sides may bind to each other with a certain strength. A tile can attach itself to a partially-formed assembly of tiles if the total binding strength between the tile and the assembly exceeds a prescribed system parameter called the temperature. The assembly process proceeds as  tiles attach themselves to the growing assembly, and stops when no more tile attachments can occur. Tile assembly is a computational process: the exposed edges of the growing crystal encode the state information of the system, and this information is modified as a new tiles attach themselves to the crystal. In fact, tile assembly can be thought of as a nondeterministic asynchronous cellular automaton, where there is a notion of a growth  starting from a seed. Tile assembly formally couples computation with shape construction, and the shape can be viewed as the final output of the tile assembly ``program". The model is capable of universal computation~\cite{winfree96win} (Turing machine simulation) and even intrinsic universality~\cite{woods2012focs} (there is a single tile set that simulates any tile assembly system). Tiles can efficiently encode shapes, in the sense that there is a close link between the Kolmogorov complexity of an arbitrary, scaled, connected shape and the number of tile types required to assemble that shape~\cite{winfree07sol}. There have been a wide selection of results on clarifying what shapes can and can not be constructed in the tile assembly model, with and without resource constraints on time and the number of required tile types (e.g.\ see surveys~\cite{patitz2012introduction,doty2012}),  showing that the tile assembly model demonstrates a wide range of computational capabilities.

There have been a number of interesting generalizations of the tile assembly model.  These include the two-handed, or hierarchical, tile assembly model where whole assemblies can bind together in a single step~\cite{Aggarwal_etal2005generalized,demaine2008staged,cannon2012two}, the geometric model where the sides of tiles have jigsaw-piece like geometries~\cite{fu2011self}, rotatable and flipable polygon and polyomino tiles~\cite{one}, and
models where temperature~\cite{Aggarwal_etal2005generalized,   kao2006reducing,summers2012temperature}, concentration~\cite{becker2010,chandran2012tile,kao2008conc,doty2010conc} or mixing stages~\cite{demaine2008staged,Staged1D_DNA2011}  are programmed. All of these models could be classed as passive in the sense that after some structure is grown, via a crystal-like growth process, the structure does not change. Active tile assembly models include the activatable tile model where tiles pass signals to each other and can change their internal `state' based on other tiles that join to the assembly~\cite{padilla2011hierarchical, padilla2012asynchronous,jonoska2012active}. This interesting generalization of the tile assembly model is essentially an asynchronous nondeterministic cellular automaton of a certain kind, and may indeed be implementable using DNA origami tiles with strand displacement signals~\cite{padilla2011hierarchical, padilla2012asynchronous}.  There are also models of algorithmic self-assembly where after a structure grows, tiles can be removed, such as the kinetic tile-assembly model~\cite{winfree98b, winfree04, chen2011optimizing} (which implements the chemical kinetics of tile assembly) and the negative-strength glue model~\cite{doty2011negative, reif2006complexity, patitz2011exact}, or the RNAse enzyme model~\cite{RNaseSODA2010, patitz2010identifying, RNAAssembly_STACS2011}. 
Although these models share properties of our model including geometry and the ability of tile assemblies to change over time, they do not share our notion of movement, something that is fundamental to our model and sets it apart from models that can be described as asynchronous nondeterministic cellular automata.

As a model of computation, stochastic chemical reaction networks are computationally universal in the presence of non-zero, but arbitrarily low, error~\cite{angluin2006fast, soloveichik2008computation}. Our model crucially uses ideas from stochastic chemical reaction networks. In particular our
update rules are all unimolecular or bimolecular (involving at most 1 or 2 monomers per rule) and our model evolves in time as a continuous time Markov process; a notion of time  we have  borrowed from the stochastic chemical reaction network model~\cite{gillespie1992rigorous}. 

Our active-self assembly model is capable of building and reconfiguring structures.  The already well-established field of reconfigurable robotics is concerned with systems composed of a large number of (usually) identical and relatively simple  robots that act cooperatively to perform a range of tasks beyond the capability of any particular special-purpose robot. Individual robots are typically capable of independent physical movement and collectively share resources such as power and computation~\cite{ murata2007self, rus2001crystalline, yim2007modular}.  One standard problem is to arrange the robots in some arbitrary initial configuration, specify some desired target configuration, and then have the robots collectively carry out the required reconfiguration from source to target.  Crystalline reconfigurable robots~\cite{rus02but, rus2001crystalline} have been studied in this way. This model consists of unit-square robots that can extend or contract arms on each of four sides and attach or detach from neighbors. For this model, Aloupis et al~\cite{Crystalline_ISAAC2008} give an algorithm for  universal reconfiguration between a pair of connected shapes that works in time merely logarithmic in shape size. As pointed out by Aloupis et al in a subsequent paper~\cite{aloupis2011efficient}, this high-speed reconfiguration can lead to strain on individual components, but they show that if each robot can displace at most a constant number of other robots, and reach at most constant velocity, then there is an optimal $\Theta(n)$ parallel time reconfiguration algorithm (which also has the desired property of working ``in-place''). Reif and Slee~\cite{reif2007optimal} also consider physicaly constrained movement, giving an optimal $\Theta(\sqrt{n})$ reconfiguration algorithm where at most constant acceleration, but up to linear speed, is permitted.  Like our nubot model, these models and algorithms implement fast parallel reconfiguration. However, here we intentionally focus on growth and reconfiguration algorithms that use very few states (typically logarithmic in assembly size) to model the fact that molecules are simple and `dumb'---however, in reconfigurable robotics it is typically the case that individual robots can store the entire desired configuration. %
Our model also has other properties that don't always make sense for macro-scale robots such as growth and shrinkage (large numbers of monomers can be created and destroyed), a presumed unlimited fuel source, asynchronous independent rule updates and Brownian motion-style agitation. Nevertheless we think it will make interesting future work to see what ideas and results from reconfigurable robotics apply to a nanoscale model and what do not.

Klavins et al~\cite{klavins04kla,lipsky04kla} model active self-assembly using conformational switching~\cite{saitou99sai} and graph grammars~\cite{ehrig79ehr}. In such work, an assembly is a graph, which can be modified by graph rewriting rules that add or delete vertices or edges. Thus the model focuses attention on the topology of assemblies. Besides permitting the assembly of static graph structures, other more dynamic structures---such as a walker subgraph that moves around on a larger graph---can be expressed. Our model also has the ability to change structure and connectivity in a way that takes inspiration from such systems, but additionally includes geometric constraints by virtue of the fact that it lives on a two-dimensional grid.   Lindenmayer systems~\cite{lindenmayer1968mathematical} are another  model where a graph-like structure is modified via insertion and addition of nodes, and where it is possible to generate beautiful movies of the growth of ferns and other plants~\cite{prusinkiewicz1991algorithmic}. Although it is indeed a model of (potentially fast) growth via insertion of nodes, it is quite different in a number of ways from our own model.

\section{Nubot model description}\label{sect:model}

  \begin{figure}[!h]
    \begin{center}
      \includegraphics[width = \linewidth]{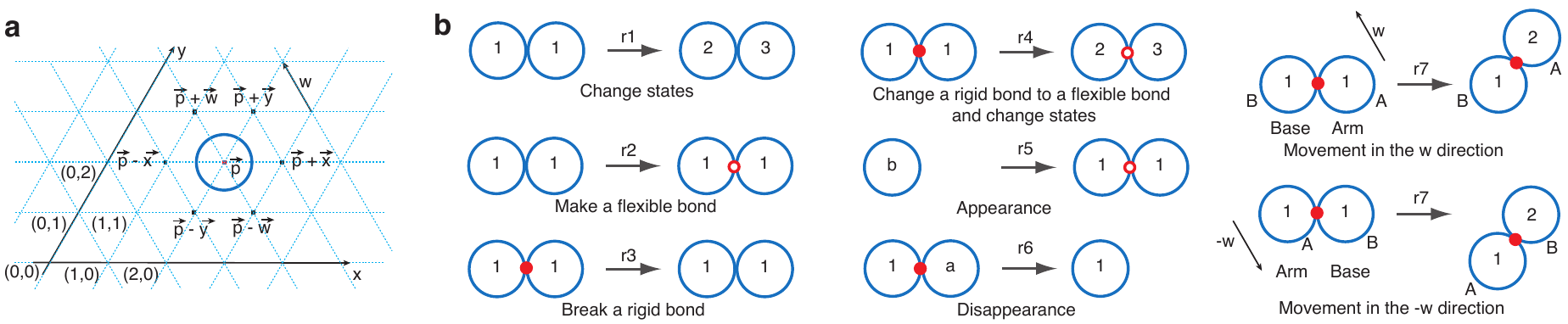}
      \caption{(a) A {\nut} in the triangular grid coordinate system. (b) Examples of {\nut} interaction rules, written formally as follows:
$r1 = (1, 1, \mathsf{null}, \vec{x})\ra(2, 3, \mathsf{null}, \vec{x}) $, 
$r2 = (1, 1, \mathsf{null}, \vec{x})\ra(1, 1, \mathsf{flexible}, \vec{x}) $,
$r3 = (1, 1, \mathsf{rigid}, \vec{x}) \ra (1, 1, \mathsf{null}, \vec{x}) $,
$r4 = (1, 1, \mathsf{rigid}, \vec{x}) \ra (2, 3, \mathsf{flexible}, \vec{x}) $,
$r5 = (b, \mathsf{empty}, \mathsf{null}, \vec{x}) \ra (1, 1, \mathsf{flexible}, \vec{x}) $,
$r6 = (1, a, \mathsf{rigid}, \vec{x}) \ra (1, \mathsf{empty}, \mathsf{null}, \vec{x}) $,
and $r7 = (1, 1, \mathsf{rigid}, \vec{x}) \ra (1, 2, \mathsf{rigid}, \vec{y}) $. 
For rule $r7$, the two potential symmetric movements are shown corresponding to two choices for arm and base, one of which is nondeterministically chosen. %
}
      \label{fig:model}
    \end{center}
  \end{figure}

This section contains the model definition. We also give a number of simple example constructions in Section~\ref{sect:examples}, which may aid the reader here.

The model uses a two-dimensional triangular grid with a coordinate system using axes $x$ and $y$ as shown in Figure~\ref{fig:model}(a). For notational convenience we define a third axis $w$, that runs through the origin and parallel to the vector $\vec{w}$ in Figure~\ref{fig:model}(a), but only axes $x$ and $y$ are used to define coordinates.  In the vector space induced by this coordinate system, the \emph{axial directions} $\mathcal{D} =  \{ \vec{w}$, $-\vec{w}$, $\vec{x}$, $-\vec{x}$, $\vec{y}$, $-\vec{y} \}$ are the unit vectors along the grid axes.  A pair $\vec{p} \in \mathbb{Z}^2$ is called a \emph{grid point} and has the set of six \emph{neighbors}  $\{ \vec{p} + \vec{u} \mid \vec{u} \in \mathcal{D} \}$. 

The basic assembly unit is called a nubot \emph{monomer}.  A monomer is a pair $X = (s_X,\vec{p}(X))$  where $s_X\in \Sigma$ is one of a finite set of states and $\vec{p}(X) \in \mathbb{Z}^2$ is a grid point. Each grid point contains zero or one {\nut}s.   
Monomers are  depicted  as state-labelled disks of unit diameter centered on a grid point.
In general, a {\em nubot} is a collection of nubot monomers.

Two neighboring monomers (i.e.\ residing on neighboring grid points) are either connected by a \emph{flexible} bond or a \emph{rigid} bond, or else have no bond between them (called a \emph{null} bond).  A flexible bond is depicted as a small empty circle and a rigid bond is depicted as a solid disk.  Flexible and rigid  bonds are described in more detail in Definition~\ref{def:move}. 

A \emph{connected component} is a maximal set of adjacent {\nuts} where every pair of {\nuts} in the set is connected by a path consisting of {\nuts} bound by either flexible or rigid bonds.  

A \emph{configuration}~$\config$ is defined to be a finite set of monomers at distinct locations and the bonds between them, and is assumed to define the state of an entire grid at some time instance.  A configuration $\config$ can be changed either by the application of an {\em interaction rule} or by a random {\em agitation}, as we now describe.

\subsection{Rules}

Two neighboring {\nuts}  can interact by an \emph{interaction rule}, $r = (s1, s2, b, \vec{u}) \rightarrow (s1', s2', b', \vec{u'})$.   %
To the left of the arrow, $s1, s2 \in \Sigma \cup \{ \mathsf{empty} \}$ are monomer states, at most one of $s1, s2$ is $ \mathsf{empty}$ (denotes lack of a monomer), 
$b \in \{\mathsf{flexible}, \mathsf{rigid}, \mathsf{null} \}$ is the bond type between them, and $\vec{u} \in \mathcal{D}$  is the relative position of the  $s2$ monomer  to the $s1$ monomer. If either of $s1,s2$ is $\mathsf{empty}$ then $b$ is $ \mathsf{null}$,  also if either or both of $s1',s2'$ is $\mathsf{empty}$ then $b'$ is $ \mathsf{null}$. The right is defined similarly, although there are some restrictions on the rules (involving $\vec{u'}$)
which are described below. The left and right hand sides of the arrow respectively represent the contents of the two monomer positions before and after the application of rule~$r$. In summary, via suitable rules, adjacent monomers can change states and bond type, one or both monomers can disappear, one can appear in an empty space, or one monomer can move relative to another by unit distance.  A rule is only applicable in the orientation specified by~$\vec{u}$, and so rules are not rotationally invariant.  The rule semantics are defined next, and a number of examples are shown in Figure~\ref{fig:model}b.

A rule may involve a movement (translation), or not. First, let us consider the case where there is no movement, that is, where $\vec{u} = \vec{u'}$. Thus we have a rule of the form $r = (s1, s2, b, \vec{u}) \rightarrow (s1', s2', b', \vec{u})$. From above, we have the restriction that at most one of $s1,s2$ is the special $\mathsf{empty}$ state (hence we  disallow spontaneous generation of monomers from completely empty space). 
{\em State change} and {\em bond change} occurs in a straightforward way and a few examples are shown in Figure~\ref{fig:model}b.
If $s_i \in \{s1,s2\}$ is $\mathsf{empty}$ and $s_i'$ is not, then the rule induces the \emph{appearance} of a new monomer. If one or both monomers go from being non-empty to being empty, the rule induces the \emph{disappearance} of monomer(s).

A \emph{movement} rule is an interaction rule where $\vec{u} \neq \vec{u'}$. More precisely, in a movement rule $d(\vec{u},\vec{u'}) = 1$, where $d(u,v)$ is Manhattan distance on the triangular grid, and none of $s1,s2,s1',s2'$ is  $\mathsf{empty}$. If we fix $\vec{u} \in \mathcal{D}$, then there are exactly two $\vec{u'} \in \mathcal{D}$ that satisfy $d(\vec{u},\vec{u'}) = 1$.  A movement  rule is applied as follows. One of the two {\nuts} is nondeterministically chosen to be the {\em base} (which remains stationary), the other is the {\em arm} (which moves). If the $s2$ monomer, denoted $X$, is chosen as the arm then $X$ moves from its current position $\vec{p}(X)$ to a new position $\vec{p}(X) - \vec{u} + \vec{u'}$. After this movement (and potential state change), $\vec{u}'$ is the relative position of the $s2'$ monomer to the $s1'$ monomer, as illustrated in Figure~\ref{fig:model}b. If the $s1$ monomer, $Y$, is chosen as the arm then $Y$ moves from  $\vec{p}(Y)$ to  $\vec{p}(Y) + \vec{u} - \vec{u'}$. Again, $\vec{u}'$ is the relative position of the $s2'$ monomer to the $s1'$ monomer.  
Bonds and states can change during the movement, as dictated by the rule. However, we are not done yet, as during a movement, the translation of the arm monomer~$A$ by a unit vector may cause the translation of a collection of {\nuts}, or may in fact be impossible; to describe this phenomenon we introduce  two definitions. 

The $\vec{v}$-boundary of a set of monomers $S$ is defined to be the set of grid locations located unit distance in the $\vec{v}$ direction from the monomers in $S$. 

\begin{definition}[Agitation set]\label{def:agitationset}
Let  $\config$ be a configuration containing  monomer $A$, and let $\vec{v} \in \mathcal{D}$ be a unit vector. The \emph{agitation set} $\mathcal{A}(\config, A, \vec{v})$ is defined to be the minimal {\nut} set in $\config$ containing $A$ that can be translated by $\vec{v}$ such that: (a)  monomer pairs in $\config$ that are joined by rigid bonds do not change their relative position to each other, (b) monomer pairs in $\config$ that are joined by flexible bonds stay within each other's neighborhood, and (c)  the $\vec{v}$-boundary of $\mathcal{A}(\config, A, \vec{v})$ contains no monomers.
\end{definition}

It can be seen that for any non-empty configuration the agitation set is always non-empty. 

Using this definition we define the {\em movable set} $\mathcal{M}(\config, A, B, \vec{v})$ for a pair of monomers $A, B$, unit vector $\vec{v}$ and configuration $C$. Essentially, the movable set is the minimal set that can be moved without disrupting existing bonds or causing collisions with other monomers.

\begin{definition}[Movable set]\label{def:move}
Let $\config$ be a configuration containing adjacent monomers $A,B$,  let $\vec{v} \in \mathcal{D}$ be a unit vector, and let~$\config'$ be the same configuration as $\config$ except that~$\config'$ omits  any bond between $A$ and $B$. The \emph{movable set} $\mathcal{M}(\config, A,B,\vec{v})$ is defined to be the agitation set $\mathcal{A}(\config', A, \vec{v})$ if $B \not\in \mathcal{A}(\config', A, \vec{v})$, and  the empty set otherwise. 
\end{definition}

\begin{figure}[t]
    \begin{center}
      \includegraphics[width = \linewidth]{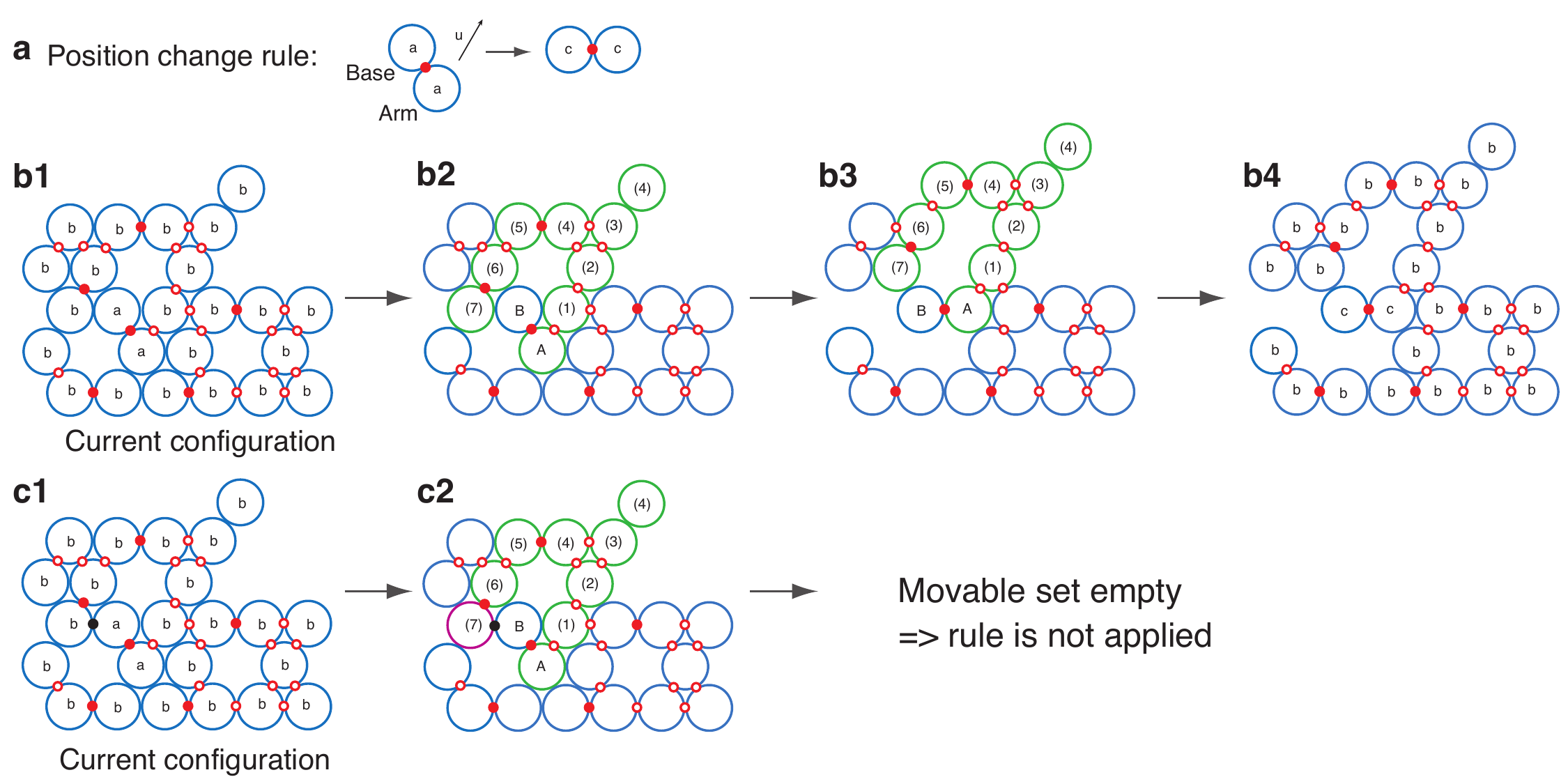}
      \caption{Two movable set examples. (a) The movement rule is $(a, a, \mathsf{rigid}, -\vec{w}) \ra (c, c, \mathsf{rigid}, \vec{x})$, where the rightmost of the two monomers is nondeterministically  chosen as the arm.
(b) Example with  a non-empty movable set. (b1)  Current configuration of the system.  (b2) Computation of the movable set. The movable set is highlighted in green, and  numbers in parentheses indicate the sequence of  incorporation of  {\nuts} into the movable set via Algorithm~\ref{alg:movable}. Monomers move as shown in~(b3). Finally, (b4) shows the configuration of the system after the application of the movement rule.
(c)~Example with an empty movable set.  (c1) A configuration that is identical to the configuration in b1, except for a single rigid bond highlighted in black. (c2)~Algorithm~\ref{alg:movable} completes at the pink {\nut}, which is blocked by the base {\nut} $B$, and hence returns an empty movable set. Thus the movement rule can not be applied.}
      \label{fig:movable}
    \end{center}
  \end{figure}

 Figure~\ref{fig:movable} illustrates this definition with two examples.

Now we are ready to define what happens upon application of a movement rule. If $\mathcal{M}(\config, A, B, \vec{v}) \neq \{ \}$, then the movement where $A$ is the arm (which should be translated  by~$\vec{v}$) and $B$ is the base (which should not be translated) is applied as follows: (1) the movable set $\mathcal{M}(\config, A, B, \vec{v})$ moves unit distance along $\vec{v}$;  (2) the states of, and the bond between, $A$ and $B$ are updated according to the rule;  (3) the states of all the  {\nuts} besides $A$ and $B$ remain unchanged and pairwise bonds remain intact (although monomer positions and bond orientations may change). 

If $\mathcal{M}(\config, A, B, \vec{v}) = \{ \}$, the movement rule is inapplicable  (the pair of monomers  $A,B$ are ``blocked'' and thus $A$ is prevented from translating).

 Section~\ref{sect:simulation} describes a greedy algorithm for computing the movable set $\mathcal{M}(\config, A, B, \vec{v})$ in time linear in assembly size. 

We note that flexible bonds are not required for any of the constructions in this paper (they are used in the construction in Figure~\ref{fig:sync} but they can be removed if we add extra monomers and rules), however we retain this aspect of the model because we anticipate that flexible bonds will be useful for future studies.

\subsection{Agitation}
 \emph{Agitation} is intended to model movement that is not a direct consequence of a rule application, but rather results from diffusion, Brownian motion,  turbulent flow or other undirected inputs of energy.  An \emph{agitation} step, applies a unit vector movement to  a monomer. This monomer then moves, possibly along with many other monomers in a way that does not break rigid nor flexible bonds. More precisely, applying a $\vec{v} \in \mathcal{D}$ agitation step to monomer~$A$ causes the agitation set $\mathcal{A}(\config, A, \vec{v})$ (Definition~\ref{def:agitationset}) to move by vector~$\vec{v}$.
During agitation, the only change in the system configuration is in the positions of the constituent {\nuts} in the diffusing component, and all the {\nut} states and bond types remain unchanged. 
 
None of the constructions in this paper exploit agitation, and all work correctly regardless of its presence or absence (due to the fact that our constructions are stable, see below). However, we feel it is important enough to be considered as part of the model definition. We leave open the possibility of designing interesting systems that exploit agitation (e.g.\ by having components interact by drifting into each other as is typical in a molecular-scale environment). 

\subsection{Stability}
The following definition is useful for proving correctness of many of our constructions.

\begin{definition}\emph{(Stable)}\label{def:stable}
A configuration $\config$   is {\em stable} if for all monomers $A$ in $\config$ and for all 6 unit vectors $\vec{v} \in \mathcal{D}$, the agitation set $\mathcal{A}(\config, A, \vec{v})$ is the entire set of monomers in $\config$. 
 \end{definition}
In other words, translating any monomer by any of the 6 unit vectors in $\mathcal{D}$ results in the translation of the entire set. This happens when monomers have a bond structure such that under agitation all monomers move together unit distance, and their relative positions remain unchanged. Hence, stable configurations have a bond structure that allows the entire structure to be ``pushed'' or ``pulled'' around by the movement of any individual monomer, without changing the relative location of any monomer to any other monomer in the configuration. Essentially, this can be used as a tool to show that a structure does not unintentionally become disconnected or end up in an unintended configuration. This is a very useful property when proving the correctness and carrying out time analysis of our constructions, and is extensively used in this paper.

\subsection{System evolution}
An \emph{assembly system} $T = (\config_0, \mathcal{R})$ is a pair  where $\config_0$ is the initial configuration, and $\mathcal{R}$ is the set of interaction rules. 
Consider two configurations $\config_i$ and $\config_j$. If $\config_j$ can be derived from $\config_i$ by a single step agitation, we write $\config_i \vdash_{{A}} \config_j$. 
 Let the relation $\vdash_{A}^*$ be the reflexive transitive closure of $\vdash_{A}$.  
 If $\config_{i} \vdash_{A}^* \config_{j}$, then $\config_{i}$ and $\config_j$ are called \emph{isomorphic} configurations, which is denoted as $\config_{i} \cong \config_{j} $. (Notice that the relative position of monomers may differ for two isomorphic configurations, although their connectivity is identical.)  
If $\config_j$ can be derived from $\config_i$ by a single application of a rule $r \in \mathcal{R}$, we write $\config_i \vdash_{r} \config_j$. 
 If $\config_i$ {\em transitions} to $\config_j$  by either a single  agitation step or a single application of some rule $r \in \mathcal{R}$, we write $\config_i \vdash_{T} \config_j$. Let the relation $\vdash_{T}^{*}$ be the reflexive transitive closure of $\vdash_T$. The set of configurations that can be \emph{produced} by an assembly system $T = (\config_0, \mathcal{R})$ is   $\mathrm{Prod}(T) = \{ \config \mid  \config_0  \vdash_{T}^* \config  \}$. The set of \emph{terminal} configurations are $\mathrm{Term}(T) = \{ \config \mid \config \in \mathrm{Prod}(T) ~\mathrm{and}~ \nexists \tilde{D} \ncong \config ~\mathrm{s.t.}~ \config \vdash_{T}^* \tilde{D} \}$. An assembly system $T$ \emph{uniquely produces} $\config$ if $\forall \tilde{D} \in \mathrm{Term}(T)$, $\config \cong \tilde{D}$.  A {\em trajectory} is a finite sequence of configurations $\config_1, \config_2, \ldots , \config_k$ where  $\config_i \vdash_T \config_{i+1}$ and $1 \leq i \leq k-1$.

An assembly system evolves as a continuous time Markov process. %
For simplicity, when counting the number of applicable rules for a configuration, a movement rule is counted twice, to account for the two symmetric choices of arm and base. If there are $k$ applicable transitions for a configuration~$C_i$ (i.e.\ $k$ is the sum of the number of rule and agitation steps that can be applied to all monomers), then the probability of any given transition being applied is $1/k$, and the time until the next transition is an exponential random variable with rate $k$ (i.e.\ the expected time is $1/k$). The rate for each rule application and agitation step is 1 in this paper (although more sophisticated rate choices can be accommodated by the model).   The probability of a trajectory is then the product of the probabilities of each of the transitions along the trajectory, and the expected time of a trajectory is the sum of the expected times of each transition in the trajectory. 
Thus, $\sum_{t \in \mathcal{T}} \mathrm{Pr}[t]  \mathrm{time}(t)$ is the expected time for the system to evolve from configuration $C_i$ to configuration $C_j$, where~$\mathcal{T}$ is the set of all trajectories from $C_i$ to  any configuration isomorphic to $C_j$, that do not pass through any other configuration isomorphic to $C_j$,  and $ \mathrm{time}(t)$ is the expected time for trajectory $t$.

The following lemma is useful for the time analysis of the  constructions in this paper.

\begin{lemma}
Given an assembly system $T$ in which all configurations in $ \mathrm{Prod}(T)$ have at most one  stable connected component, 
the expected time from configuration $C_i$ to configuration $C_j$ is the same with or without agitation steps.
\end{lemma}
\begin{Pf}
Assembly system $T$ has a corresponding Markov process $M$. We consider another Markov process $M'$. Each state $S_C$ in $M'$ corresponds to all configurations in $ \mathrm{Prod}(T)$ that are isomorphic to a  particular configuration $C$, which by hypothesis are all stable. For each transition in $M$ from configuration $C_x$ to $C_y$, there is also a transition from $S_{C_x}$ to $S_{C_y}$ with the same transition rate. The expected time from configuration $C_i$ to configuration~$C_j$ in the assembly system $T$ is the same as the expected time from state~$S_{C_i}$ to first reach~$S_{C_j}$ in Markov process $M'$ since every trajectory in $M$ corresponds to a trajectory in $M'$. Furthermore, all agitation steps in $T$ corresponds to self-loops in $M'$. Therefore, removing the agitation steps in $T$ is equivalent to removing some self-loops in $M'$ and does not change the expected time to transition from one state to another.
\end{Pf}

Since all of our constructions satisfy the assumption that all producible configurations  have only a single connected stable component, we ignore agitation steps in the time analysis of the  constructions in this paper.  

\section{Examples}\label{sect:examples}
\begin{figure}[h] %
    \begin{center}
      \includegraphics[width = \linewidth]{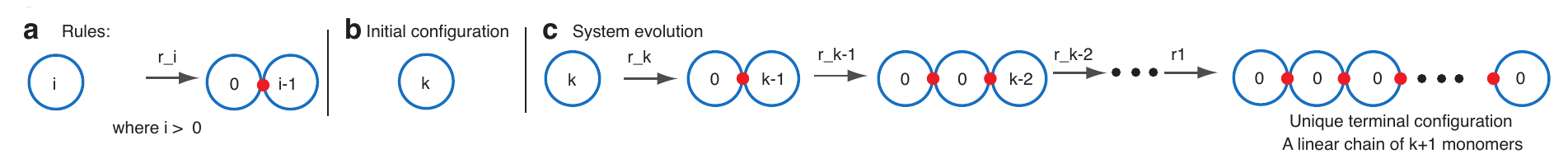}
      \caption{Growth of a linear chain.
(a) Rule set: $\mathcal{R}_k = \{ r_i \mid r_i = (i, \mathsf{empty}, \mathsf{null}, \vec{x}) \ra (0, i-1, \mathsf{rigid}, \vec{x}), \mathrm{~where~}  0 < i \leq k\}$.
(b) Initial configuration of the system. 
(c) Evolution of a system trajectory over time.}
      \label{fig:c-growth}
    \end{center}
  \end{figure}

\begin{figure}[h]
    \begin{center}
      \includegraphics[width = \linewidth]{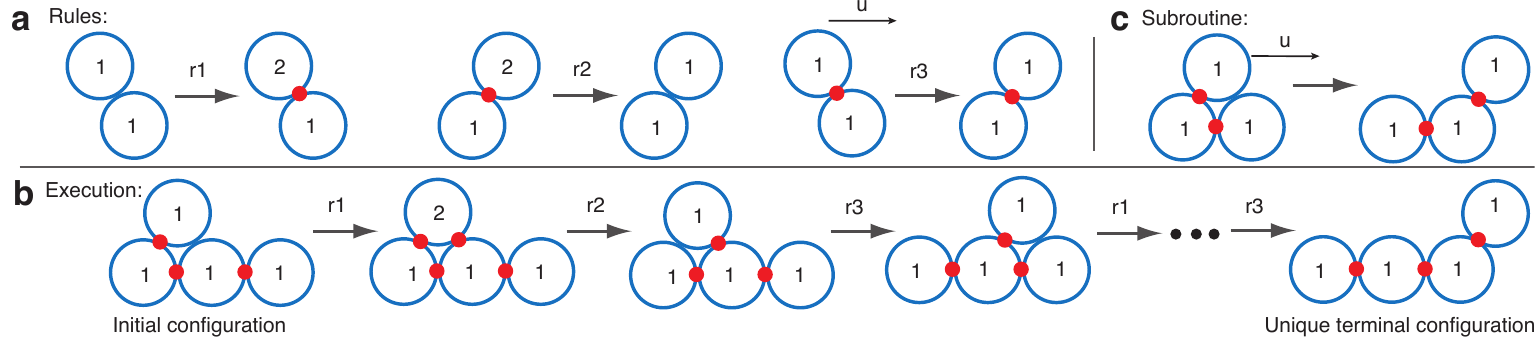}
      \caption{Autonomous unidirectional motion of a walker along a linear track. (a) Rule set:
$r1 = (1, 1, \mathsf{null}, -\vec{w}) \ra (2, 1, \mathsf{rigid}, -\vec{w})$,
$r2 = (1, 2, \mathsf{rigid}, \vec{y}) \ra (1, 1, \mathsf{null}, \vec{y})$, and
$r3 = (1, 1, \mathsf{rigid}, \vec{w}) \ra (1, 1, \mathsf{rigid}, \vec{y})$. (b) Evolution of an example initial configuration. (c)~A subroutine abstraction.
}
      \label{fig:c-walking}
    \end{center}
  \end{figure}

\begin{figure}[!h] %
    \begin{center}
      \includegraphics[width = \linewidth]{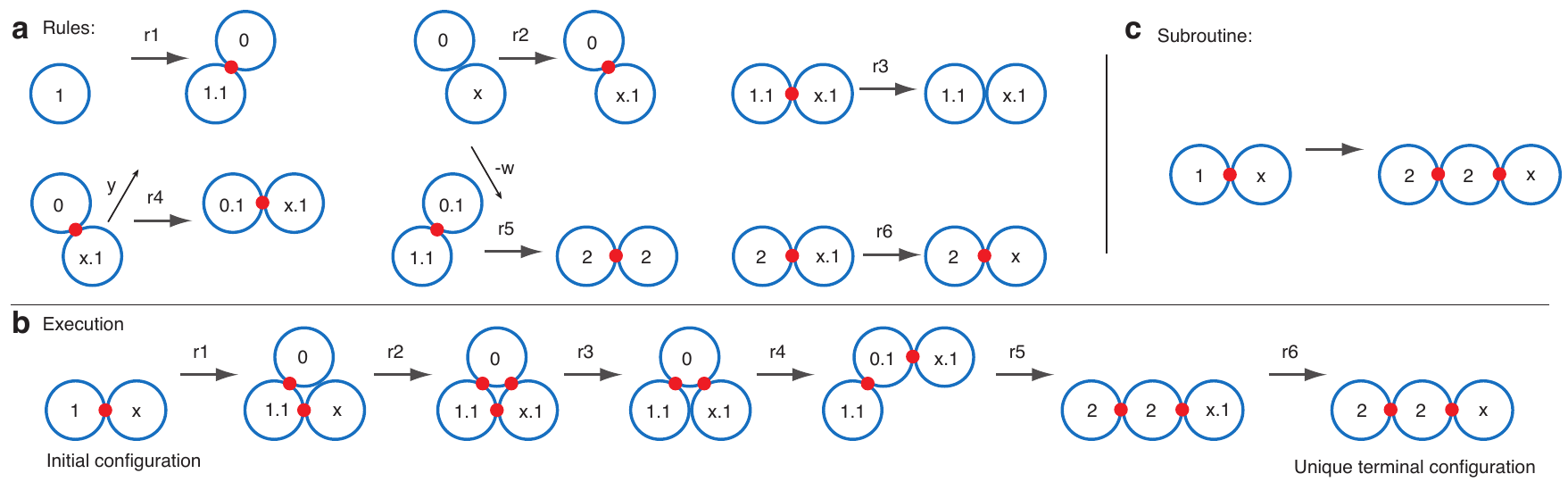}
      \caption{Insertion of a single monomer between two others.
(a) Rule set: 
$r1 = (1, \mathsf{empty}, \mathsf{null}, \vec{y}) \ra (1.1, 0, \mathsf{rigid}, \vec{y})$,
$r2 = (0, x, \mathsf{null}, -\vec{w}) \ra (0, x.1, \mathsf{rigid}, -\vec{w})$,
$r3 = (1.1, x.1, \mathsf{rigid}, \vec{x}) \ra (1.1, x.1, \mathsf{null}, \vec{x})$,
$r4 = (0, x.1, \mathsf{rigid}, -\vec{w}) \ra (0.1, x.1, \mathsf{rigid}, \vec{x})$,
$r5 = (1.1, 0.1, \mathsf{rigid}, \vec{y}) \ra (2, 2, \mathsf{rigid}, \vec{x})$, and
$r6 = (2, x.1, \mathsf{rigid}, \vec{x}) \ra (2, x, \mathsf{rigid}, \vec{x})$.
(b) Evolution of an example initial configuration.  (c)~A subroutine abstraction.}
      \label{fig:c-insertion}
    \end{center}
  \end{figure}

\begin{figure}[h]
    \begin{center}
      \includegraphics[width = \linewidth]{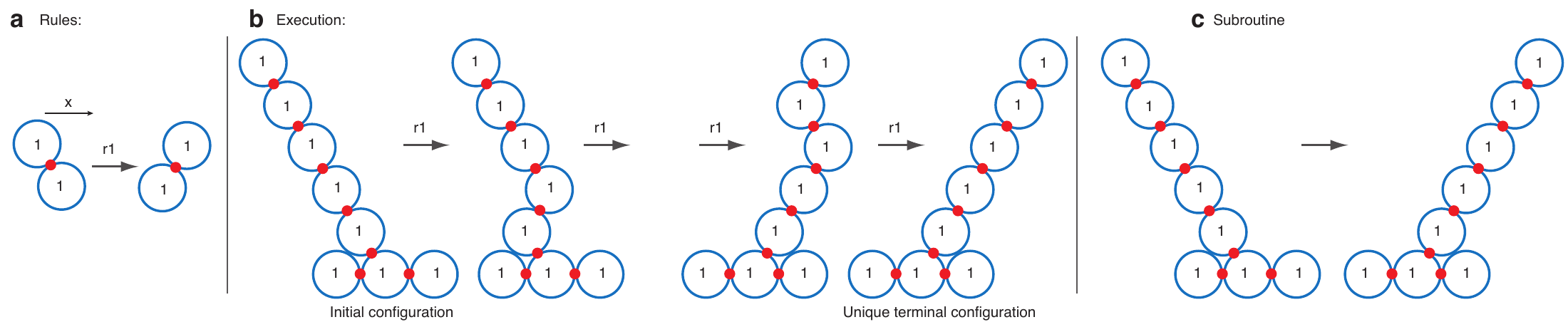}
      \caption{Rotation of a long arm.
(a) Rule set: 
$r1 = (1, 1, \mathsf{rigid}, \vec{w}) \ra (1, 1, \mathsf{rigid}, \vec{y})$.
(b) Evolution of an example initial configuration to a terminal configuration. 
(c)~A subroutine abstraction.}
      \label{fig:rotation}
    \end{center}
  \end{figure}

\begin{figure}[h]
    \begin{center}
      \includegraphics[width = 2.7in]{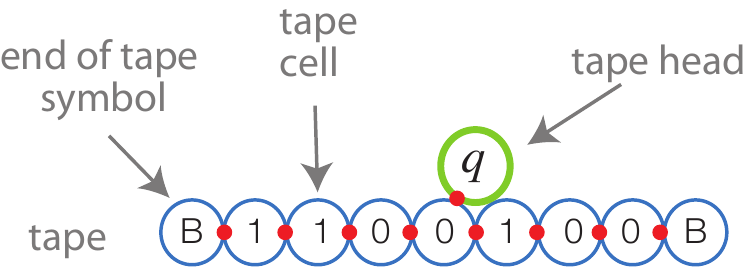}
      \caption{Turing machine example. The Turing machine program is stored in the rule set. New tape monomers can be created as needed. \dw{ Make TM fig more interesting (show a TM program and tape with arrow going to nubots rules and line of monomers.): add in program fragment, right-arrow, rule-set. }}
      \label{fig:TM-example}
    \end{center}
  \end{figure}

Figures~\ref{fig:c-growth} to~\ref{fig:TM-example} show a number of examples. 
Figure~\ref{fig:c-growth} depicts a simple assembly system that grows a line of monomers. Figure~\ref{fig:c-growth}b shows the initial configuration $\mathcal{C}_0$, which consists of a single monomer with state $k  \in \mathbb{N}$. This system evolves as shown in Figure~\ref{fig:c-growth}c, and uniquely produces a linear chain with $k+1$ monomers, with each monomer having a final state of~$0$. Since the system involves $k$ \emph{sequential} events each with unit expected time, it takes expected time $k$ to complete.

Figure~\ref{fig:c-walking} depicts the autonomous unidirectional motion of a single-monomer ``walker" along a linear track of monomers. It takes $O(n)$ expected time for a {\nut} to move $n$ steps. The rules depicted in Figure~\ref{fig:c-walking}a implement the subroutine depicted in Figure~\ref{fig:c-walking}c. 

Figure~\ref{fig:c-insertion} describes insertion of a single monomer between two other monomers, which occurs in constant expected time.

Figure~\ref{fig:rotation} describes the rotation of a long arm. As the motion of each monomer is independent of the motion of the other monomers, it takes $O(\log n)$ expected time to complete the rotation of an arm of $n$ monomers.  Interestingly, this example captures a kind of movement and speed that is impossible to achieve with cellular automata, but has an analog  in reconfigurable robotics~\cite{Crystalline_ISAAC2008}.

Figure~\ref{fig:TM-example} shows a simple Turing machine example. The Turing machine program is stored in the rule set and directs the  tape head monomer to walk left or right, and to update the relevant tape monomer. The Turing machine state is stored in the tape monomer, currently the Turing machine is in state $q$. If the Turing machine requires a longer tape, new tape cell monomers are created to the left or right of the end of tape marker $B$. This example shows that the model is capable of algorithmically directed behavior.

\section{System simulation}\label{sect:simulation}
In this section we show that there is an algorithm that simulates one of the trajectories of an assembly system in time $O(tn^2)$ where $t$ is the number of configuration transitions in the trajectory and $n$ is the maximum number of monomers of any configuration in the trajectory. This shows that simulation of individual trajectories is tractable in terms of the number of rule applications, and in terms of the number of monomers.\footnote{This forms the basis of our software simulator for the model.}\nadine{link to simulations?} In fact, it is not difficult to imagine other variations on the model where simulation of a single, non-local, rule is an intractable problem.  The existence of our algorithm gives evidence that our rules, in particular the movement rule, are in some sense reasonable. %

\subsection{Simulation of a single step} 
The continuous-time Markov process that describes a  trajectory is simulated using a discrete time-algorithm.  Essentially, the algorithm examines the grid contents and using a local neighborhood of radius 2, a list of potentially applicable rules are generated. The list also contains $6n$ potentially applicable agitation steps (6 directions for each of the $n$ monomers). All of this can be done in $O(n)$ time. The algorithm then, uniformly at random, picks an event from the list to apply, if the rule or agitation step can indeed be applied  the grid contents are updated accordingly. 
Besides the movement rules and agitation steps, the other rule types can be easily simulated in time $O(n)$, since at most two grid sites are affected. 
As described below, movement is simulated  in time $O(n^2)$. Hence a single step is simulated in time $O(n^2)$.

Due to its non-local nature, the movement rule is the most complicated  rule type to simulate. Algorithm~\ref{alg:movable} below calculates the moveable set for a given movement rule in  time $O(n)$. We may have to try this algorithm $< n$ times  before we can find a non-empty movable set and can apply a movement rule, or decide that there is no applicable movement rule.  Applying a rule simply involves translating $< n$ monomers by unit distance, which can be done in $O(n)$ time.  Hence movement can be simulated in $O(n^2)$ time. Agitation is simulated similarly.

\subsection{Computing the movable set}\label{sect:simulation-movable}
We  describe a greedy algorithm for computing the movable set $\mathcal{M}(\config, A, B, \vec{v})$ of monomers for a movement rule where $\config$ is a configuration, $A$ is an arm monomer, $B$ is a base monomer and $\vec{v}\in \mathcal{D}$ the  unit vector describing the translation of $A$. The algorithm takes time linear in the number of monomers. Figure~\ref{fig:movable} shows two examples of computing the movable set.

\begin{algorithm}\label{alg:movable} Compute movable set $\mathcal{M}(\config, A, B, \vec{v})$. 
\begin{itemize}
       \item Step 1.  Let $\mathcal{M} \leftarrow \{ A \}$, $\mathcal{F} \leftarrow \{ A \}$, $\mathcal{B} \leftarrow \{ \}$.
       \item Step 2. Compute the \emph{blocking set} $\mathcal{B}$ for the {\em frontier set} $\mathcal{F}$ along $\vec{v}$, as follows. \\
         For each {\nut} $X \in \mathcal{F}$, do:
	\begin{enumerate}
		\item If $\vec{p}(X) + \vec{v}$ is occupied by $Y \notin \mathcal{M}$, then $\mathcal{B} = \mathcal{B} \cup \{ Y \}$;
		\item If $X$ is bonded to $Y \notin \mathcal{M}$, and if translating $X$ by $\vec{v}$ without translating $Y$ would disrupt the bond between $X$ and $Y$,  then $\mathcal{B} = \mathcal{B} \cup \{ Y \}$; (Ignore the special case where $X = A, Y = B$).
	\end{enumerate}       
       \item Step 3. Inspect the blocking set:
       \begin{enumerate}
       	\item If $B \in \mathcal{B} $, return $\{ \}$;
	\item If $\mathcal{B} = \{\}$, return $\mathcal{M}$;
	\item Otherwise, let $\mathcal{M} \leftarrow \mathcal{M} \cup \mathcal{B}$, $\mathcal{F} \leftarrow \mathcal{B}$, $\mathcal{B} \leftarrow \{\}$, and go to Step 2.
       \end{enumerate}
\end{itemize}
\end{algorithm}

\begin{lemma}
Algorithm~\ref{alg:movable} identifies the movable set $\mathcal{M}(\config, A, B, \vec{v})$ in time $O(n)$, where $n$ is the number of {\nuts} in $\config$.
\end{lemma}
\begin{Pf}
To argue that the algorithm identifies the movable set we consider two cases: the algorithm completes with (1) a non-empty set, and (2) an empty set. 

In Case (1) when the algorithm completes with a non-empty set $\mathcal{M}$, we only need to prove the following claims: (1.1)  $\mathcal{M}$ contains $A$ but not $B$, (1.2) $\mathcal{M}$ can move  unit distance along $\vec{v}$ without causing {\nut} collision nor bond disruption, and (1.3) $\mathcal{M}$ is the minimal set that satisfies (1.1) and (1.2).  

Claim (1.1) follows directly from step 1 and step 3.1. 

To prove Claim (1.2), assume, for contradiction, that there exists a {\nut} $Y \notin \mathcal{M}$ that blocks a {\nut} $X \in \mathcal{M}$. Therefore, when $X$ gets first incorporated into $\mathcal{M}$, $Y$ must be incorporated into $\mathcal{M}$ in the next round of execution of step 2. This contradicts $Y \notin \mathcal{M}$. Therefore, Claim (1.2) must be true. 

To prove Claim (1.3), assume, for contradiction, that there exists a set $\mathcal{C} \subset \mathcal{M}$, $A \in \mathcal{C}$ such that  $\mathcal{C}$ can move by $\vec{v}$. The first {\nut} in $\mathcal{M} \setminus \mathcal{C}$ that gets incorporated in $\mathcal{M}$ must block some monomer in $\mathcal{C}$, which contradicts that $\mathcal{C}$ can move a unit distance along $\vec{v}$. Therefore, Claim (1.3) holds. 
  
In Case (2), we know from Claim (1.3) that any movable set that contains $A$ must contain every {\nut} in $\mathcal{M}$ and thus contain $B$. Therefore, the movable set must be empty.
\end{Pf} 

We note that the nondeterministic choice of which monomer is the arm and which is the base can make a difference in the resulting configuration: for example switching the arm and base {\nuts} in Figure~\ref{fig:movable}a  will induce a different movable set. In this paper we do not exploit such asymmetric nondeterministic choices.

\section{Efficient growth of simple shapes: lines and squares}\label{sect:linesSquares}

In this section, we show how to efficiently construct lines and squares in time and number of monomer types logarithmic in shape size.  We give a fast  (logarithmic time) method to synchronize a line of monomers: the procedure  detects in logarithmic time in line length whether all monomers in the line are in the same state. 
We also give a Chernoff bound lemma that  aids in the time-analysis of these and other systems. %

\subsection{Line}\label{sect:chain}
\begin{theorem}\label{thm:line}
A line of monomers of length $n \in \mathbb{N}$ can be uniquely produced in expected time $O(\log n )$ and with $O(\log n)$ states. 
\end{theorem}
\begin{Pf}  We first describe the construction, then prove correctness and conclude with a time analysis.

Description: To build a line of length $n$, from the start monomer $s.n$, we first (sequentially) generate a short line of $p = O(\log n)$ monomers with respective states $K = \{ k_1 ,k_2,\ldots,k_p \}$, where $\sum_{k \in K} 2^k = n$. Figure~\ref{fig:t-chain2} illustrates this first step.  Then, each monomer with state $k \in K$  efficiently builds a line of length $2^{k}$ as described below. %

  \begin{figure}[t]
    \begin{center}
      \includegraphics[width = \linewidth]{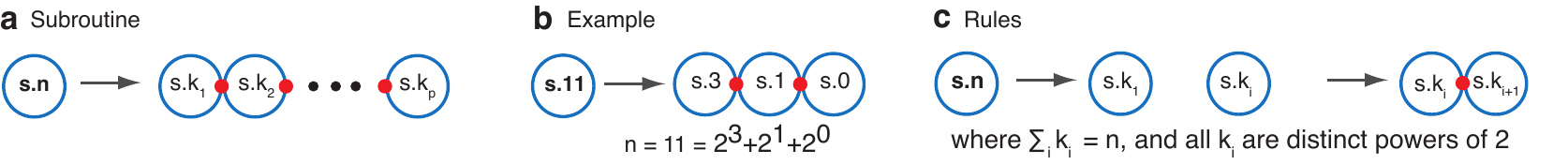}
      \caption{Building a line of length $n\in \mathbb{N}$ by decomposing into $O(\log n)$ lines whose lengths are distinct powers of 2.}
      \label{fig:t-chain2}
    \end{center}
  \end{figure}

  \begin{figure}[p]
    \begin{center}
      \includegraphics[width = 1\linewidth]{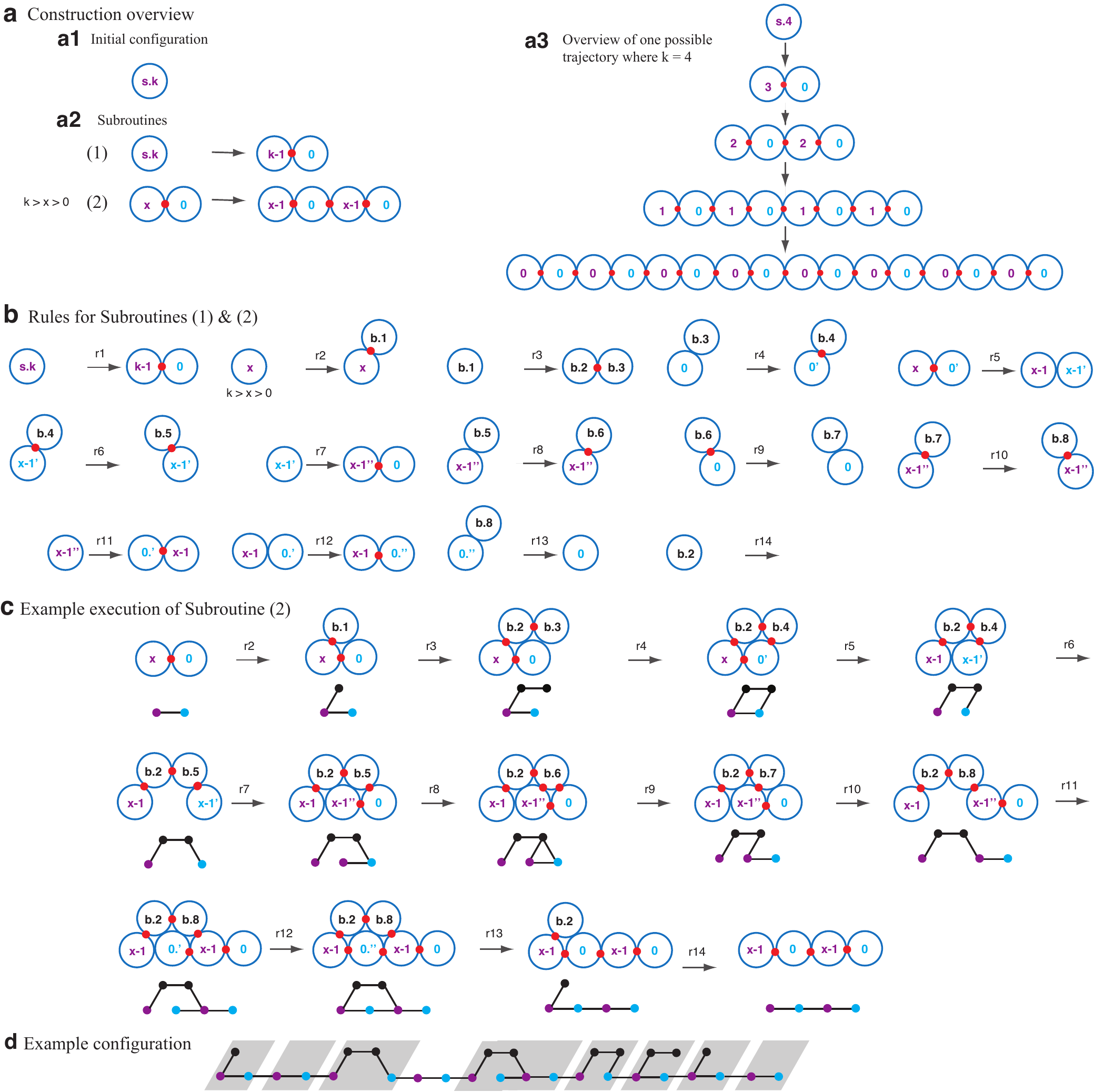} %
      \caption{Construction that builds a length $2^k$ line in expected time $O(k)$. (a)~ Overview: 1 monomer in state $x \in \mathbb{N}$  creates 2 in state $x-1$, and this  happens independently in parallel along the entire line as it is growing. The blue/purple colors are for readability purposes only. %
       (b)~Rules for subroutines (1) \& (2), for all $x$ where $k > x > 0$. %
       (c)~Example execution of 13 steps, starting with a left-right pair of monomers. The ``stick and dot'' illustrations emphasize bond structure,   representing bonds and monomers respectively. 
       (d)~Example configuration, in stick and dot notation, that emphasizes parallel asynchronous rule applications. Each gray box shows the application of a subroutine.}
      \label{fig:line}
    \end{center}
  \end{figure}

 Figure~\ref{fig:line} gives an overview of the main construction, as well as many of the rules. The  idea is to quickly build a line of length $2^k$, by having the start monomer, $s.k$,  create 2 monomers (one of which is in state $k-1$), which in turn create 4 monomers (2 of which are in state $k-2$), and so on until there are $2^k$ monomers with state 0. An overview of a possible trajectory of the system is given in Figure~\ref{fig:line}a3. However, as the model is asynchronous, most trajectories are not of this simple form.

The construction can be described using the two subroutines shown in Figure~\ref{fig:line}a2.  Subroutine (1) consists of a single rule that is applied only once, to the seed monomer.  Subroutine  (2) consists of $k-1$ sets of rules, one for each $x$ where $k > x >0$. A schematic  of one of these $k-1$ rule sets is given in Figure~\ref{fig:line}b, with an example execution of Subroutines (1) and (2) in Figure~\ref{fig:line}c. %

Subroutine (2) begins with a single pair of monomers with states $x, 0$ and ends with four monomers in states $ x-1, 0, x-1, 0$. Monomers are shown as left (purple), right (blue) pairs to aid readability. The rules for Subroutine (2) are given in Figure~\ref{fig:line}b and an example can be seen   in Figure~\ref{fig:line}c. The subroutine works as follows. Each monomer on the line has a left/right component to its state: left is colored purple, right is colored  blue. The initial $x_{\mathrm{left}}, 0_{\mathrm{right}}$ monomers send themselves to state $x-1_{\mathrm{left}},0_{\mathrm{right}}$ while inserting two new monomers to give the pattern  $x-1_{\mathrm{left}},0_{\mathrm{right}}, x-1_{\mathrm{left}}, 0_{\mathrm{right}}$, as indicated in Subroutine (2). To achieve this, the initial pair of monomers create a ``bridge'' of 2  monomers on top, and by using movement and appearance rules two new monomers are inserted. The bridge monomers are then deleted and we are left with four monomers. Throughout execution, all monomers are connected by rigid bonds so the entire structure is {\em stable}. Subroutine (2) completes in constant expected time 13.

 Subroutine (2) has the following properties: (i) during the application of its rules to an initial pair of monomers $x_{\mathrm{left}}, 0_{\mathrm{right}}$  it does not interact with any monomers outside of this pair, and (ii) a left-right pair creates two adjacent left-right pairs. Intuitively, these properties imply that along a partially formed line, multiple subroutines can  execute asynchronously and in parallel, on disjoint left-right pairs, without interfering with each other. 

Correctness: We argue by induction that the line completes with $2^k$ monomers in state $0$. %
The initial rule (Subroutine (1) in Figure~\ref{fig:line}a2) creates a left-right pair of monomers $k-1_{\mathrm{left}},0_{\mathrm{right}}$; the base case. 
For the inductive case, consider an arbitrary, even length, line of monomers $\ell_{j}$  where all monomers are arranged in left-right pairs of the form $x_{\mathrm{left}},0_{\mathrm{right}}$, where either $x \in \mathbb{N}$. For any left-right pairs of the form $0_{\mathrm{left}},0_{\mathrm{right}}$, no rules are applicable (they have reached their final state).  
For all other left-right pairs $x_{\mathrm{left}},0_{\mathrm{right}}$ Subroutine (2) is applicable. Choose any such left-right pair, and consider the new line $\ell_{j+1}$ created after applying Subroutine (2). The new line $\ell_{j+1}$ is identical to $\ell_{j}$ except that our chosen pair $x_{\mathrm{left}},0_{\mathrm{right}}$ has been replaced by $x-1_{\mathrm{left}}, 0_{\mathrm{right}},x-1_{\mathrm{left}},0_{\mathrm{right}}$. Line  $\ell_{j+1}$ shares the following property with line $\ell_{j}$: all monomers are in left-right pairs. Hence, except for (already completed) $0_{\mathrm{left}},0_{\mathrm{right}}$  pairs,  Subroutine (2)  is applicable to every left-right pair of $\ell_{j+1}$, and so by induction we maintain the property that rules can be correctly applied. Furthermore,  application of Subroutine (2) leaves one 0 state untouched, creates a new 0 state, and creates two new $x-1$ states. Hence, eventually we get a line where  all states are 0 and no rules are applicable. The fact that the line  grown from the monomer with state $s.k$ has length $2^k$ follows from a straightforward counting argument.

Time analysis: Consider any pair of adjacent monomers $0_{\mathrm{left}},0_{\mathrm{right}}$ in a final line of length $2^k$. The number of rule applications from the start monomer $s.k$ to this pair is $k$ (i.e.\ using Figure~\ref{fig:line}a2, apply Subroutine (2) $k$ times to get this pair). Given that these $k$ rule applications are applied independently  (without interference from other rules that are acting on other monomers on the line) and in sequence the expected time to generate our chosen pair is $O(k)$. There are $2^{k-1}$ of these $0_{\mathrm{left}},0_{\mathrm{right}}$ pairs in a final line of length $2^k$, giving an $O( k \log 2^{k-1}) = O(k^2)$ bound on the expected time for the line to finish. In a line of length $n \in \mathbb{N}$   we have $O(\log n)$  lines, each of length a power of 2, being generated in parallel (using the technique in Figure~\ref{fig:t-chain2}), giving an  expected time of $O(\log^2 n \log \log n)$ for the length~$n$ line to complete. This analysis  can be improved using Chernoff bounds. Specifically, in  Lemma~\ref{lem:chernoff}  we  choose  $m = n/2$ and $a_1,a_2,\ldots , a_m$ to be the $n/2$ rule applications that generate the $n/2$ pairs of $0_{\mathrm{left}},0_{\mathrm{right}}$ monomers in the final configuration. Each $a_i$ requires $ k \leq 2 \log_2 n $ insertions to happen before it.  Therefore, the expected time for the line to finish is $O(k)=O(\log n)$.

Number of states: Subroutine (1) has 1 rule. Subroutine (2) has $O(1)$ states for each $x \in {1,\ldots,k-1}$, hence the total number of states is   $O(k)$. \end{Pf}

\begin{lemma}\label{lem:chernoff}
In an assembly system, if there are $m$ rule applications $a_1, a_2,\dots, a_m$ that must happen, and 
\begin{enumerate}
\item the desired configuration is reached as soon as all $m$ rule applications happen, 
\item for any specific rule application $a_i$ among those $m$ rule applications, there exist at most $k$ rule applications $r_1, r_2, \dots, r_k$ such that $a_i=r_k$ and for all $j$, $r_j$ can be applied directly after $r_1, r_2, \dots, r_{j-1}$ have been applied, regardless of whether other rule applications have happened or not,
\item $m\leq c^k$ for some constant $c$, 
\end{enumerate}
then the expected time to reach the desired configuration is $O(k)$.
\end{lemma}
\begin{Pf}
From the assumptions, the time $T_i$ at which the rule application $a_i$ happens is upper bounded by the sum of $k$ mutually independent exponential variables, each with mean $1$ for every~$k$. Using Chernoff bounds for exponential variables~\cite{motwani}, it follows that 
\[
\mbox{Prob}[ T_i > k(1 + \delta)] \leq \big( \frac{1+\delta}{e^\delta} \big)^k.
\]
Let $T$ be the time to first reach the desired configuration. From the union bound, we know that 
\[
\mbox{Prob}[T > k(1+\delta)] \leq m\big( \frac{1+\delta}{e^\delta} \big)^k \leq \big( \frac{c(1+\delta)}{e^\delta} \big)^k \leq c^{k}e^{-\frac{k\delta}{2}}, \mbox{for all $\delta \geq 3$}.
\]
Therefore, the expected time is $E[T] = O(k)$.
\end{Pf}

For some constructions it is useful to have a procedure to efficiently (in  time and states logarithmic in $n$) detect when a large number ($n$) of monomers are in a certain state. Here we give such a procedure. Specifically, we show that after growing a line, we can use a fast signaling mechanism to synchronize the states of all the monomers in the completed line.

\begin{theorem}[Synchronized line]\label{thm:sync-dw}
In time $O(\log n)$, with $O(\log n)$ states, a line of length $n \in \mathbb{N}$ can be uniquely produced in such a way that each {\nut}  switches to a prescribed final state, but only after all insertions on the line have finished. 
\end{theorem}
\begin{Pf}
The basic idea is to build a {\em synchronization row} of monomers below the line. This row is grown only in regions of the line that have finished growth (are in state~0), and takes time $O(\log n)$ time to grow.  The bond structure of the synchronization row is such that  upon the completion of the entire line, a relative shift of the synchronization row to the line occurs (in $O(1)$ time), informing the line monomers to switch to a prescribed final state. Finally, the synchronization row is deleted, in $O(\log n)$ time.

\begin{figure}[t]
\begin{center}
\includegraphics[width = \linewidth]{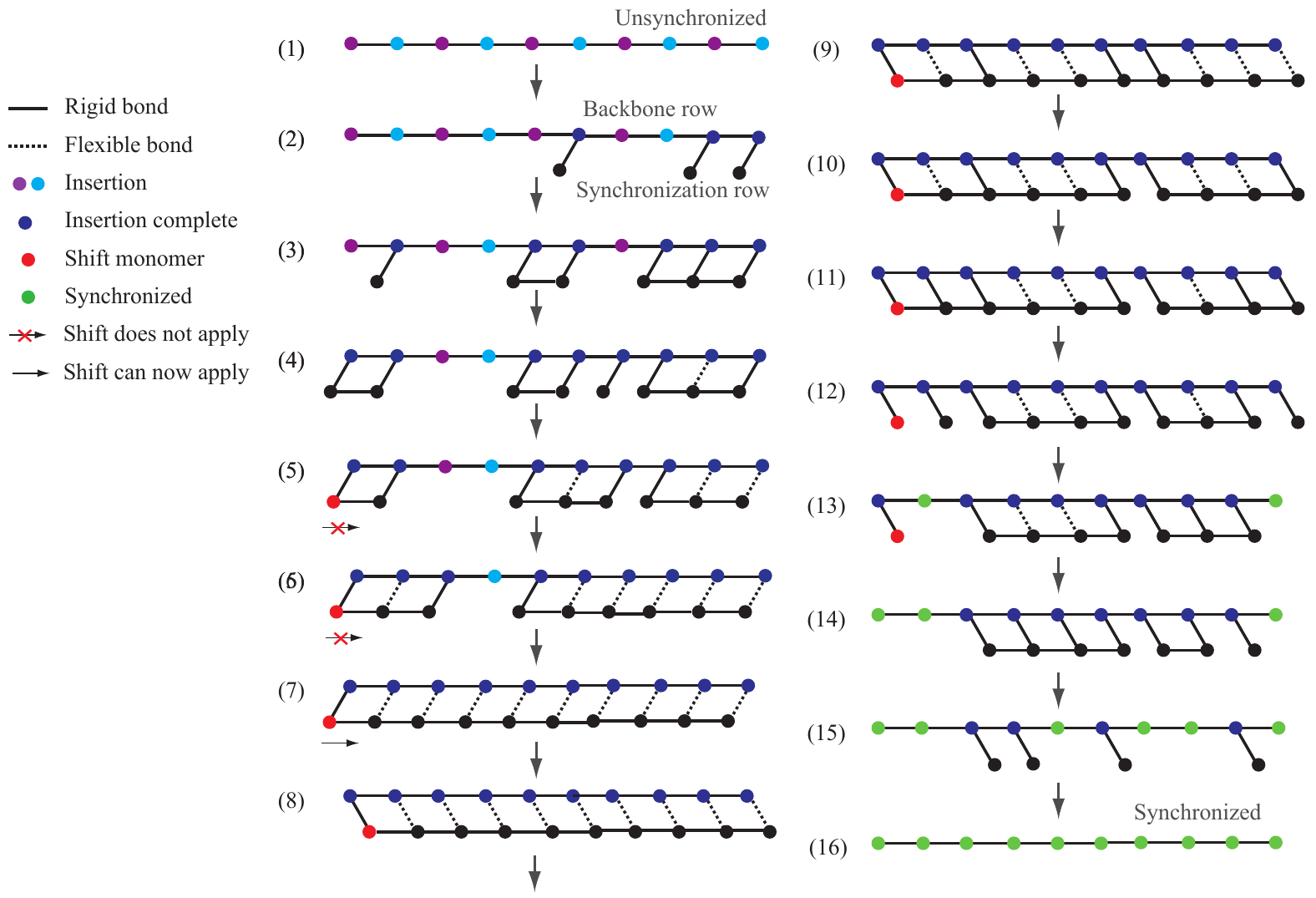}
\caption{Synchronization mechanism for Theorem~\ref{thm:sync-dw} that quickly, in $O(\log n)$ expected time, sends a signal to $n$ monomers in a line. Stick and dot notation is used to emphasize the bond structure throughout. A single movement, or shift, between configurations (7) and (8) sends the signal to all monomers. The structure maintains stability throughout execution. }
\label{fig:sync}
 \end{center}
\end{figure}

Figure~\ref{fig:sync} describes the synchronization mechanism. Starting from a seed monomer we grow a line (1), as each monomer on the line reaches state $0$ (and so has finished inserting) it grows a {\em synchronization} monomer (in black) below it (2), joined to the line with a rigid bond.  In the following we want to always ensure that the entire structure is {\em stable}. Neighboring synchronization monomers form horizontal rigid bonds (3). Any synchronization monomer that is joined with its two horizontal neighbors changes its bond to the line from rigid to flexible (the rightmost monomer is a special case, it changes to flexible immediately upon bonding with a horizontal neighbor). According to this rule we eventually get to configuration (7) where the entire synchronization row  is bonded to the backbone row by flexible bonds, except for the leftmost pair. At this time, the leftmost pair is, for the first time, able to execute a movement rule (8) which shifts the synchronization row to the right, relative to the backbone row. Backbone monomers can detect this shift.

From here on the aim is to delete the synchronization row, while maintaining the property of stability. This is achieved in a manner inversely analogous to before: synchronization monomers create rigid bonds with the backbone, then delete their bonds to their horizontal neighbors only when the neighbors have formed vertical rigid bonds.

From Theorem~\ref{thm:line} the expected time to grow the unsynchronized line, and to then grow the additional synchronization row is $O(\log n)$ (the addition of the synchronization row requires $O(1)$ for each monomer in the unsynchronized line). The movement rule that underlies the synchronization then takes expected time $O(1)$, and a further $O(\log n)$ expected time is required to delete the length $n$ synchronization row. the number of states is dominated by the $O(\log n)$ states to build a line (Theorem~\ref{thm:line}), as the synchronization mechanism itself can be executed using $O(1)$ states (Figure~\ref{fig:sync}).  \end{Pf}

\subsection{Square}\label{sect:square}

Square building is a common benchmark problem in self-assembly.
  \begin{figure}[h]
    \begin{center}
      \includegraphics[width = \linewidth]{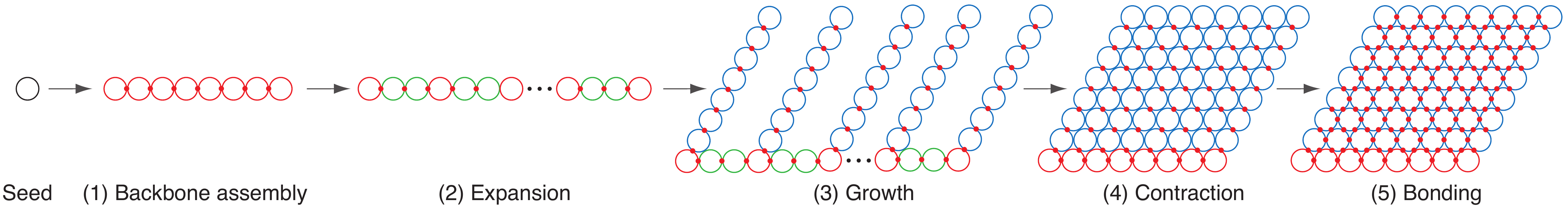}
      \caption{Building a square in $O(\log n)$ time, using $O(\log n)$  states.}
      \label{fig:t-square}
    \end{center}
  \end{figure}

\begin{theorem}\label{theorem:square}
An $n \times n$ square can be constructed in time $O(\log n)$ using $O(\log n)$ states.
\end{theorem}

\begin{Pf}
Figure~\ref{fig:t-square} contains an overview. Using the construction in Theorem~\ref{thm:sync-dw},  we first assemble a horizontal {\em backbone} line of length $n$. %
When a given monomer pair on the line finishes insertion (i.e.\ reach states $0,0$), the line then expands by a factor of 3 at that location. Every third monomer in the expanded backbone grows a vertical line of length $n$, the previous expansion ensures that each vertical line has sufficient space to grow. Each vertical line synchronizes upon completion. This synchronization signals to the backbone line to contract by factor of 3, essentially bringing all $n$ vertical lines into contact. Adjacent vertical lines form rigid bonds, so that the final shape is fully connected.

To analyze the expected time to completion we first consider the expected time for the $n$ vertical rows to be a situation where all synchronization rows have grown and they are about to apply the synchronization steps. Each synchronization monomer can grow independently of all others, and depends only on $\leq 4 \lceil \log_2 n \rceil$ prior insertion events to happen. By setting $m = n^2 - n$ in Lemma~\ref{lem:chernoff} and $k = O( \log n )$ the expected time is $O(\log n)$. The synchronizations then apply in expected time $O(\log n)$, as do the final folding and bonding steps.  The number of states to build and synchronize each line is $O(\log n)$, a constant number of other states are used in the construction. 
\end{Pf}

\section{Computable shapes}\label{sect:complexityOne}

Let $|n| = \lceil \log_2 n \rceil$ be the length of binary string encoding $n \in \mathbb{N}$.
\begin{theorem}\label{thm:shape}
An arbitrary connected computable 2D shape of size $\leq \sqrt{n} \times \sqrt{n}$ can be  constructed in expected time $O(\log^2 n + t(|n|)) $  using  $O(s + \log n)$ states. Here, $t(|n|)$ is the time required for a program-size $s$ Turing machine to compute, given the index of a pixel $n$, whether the pixel is present in the shape. 
\end{theorem}

The remainder of this section contains the proof.

  \begin{figure}[p]
    \begin{center}
      \includegraphics[width = \linewidth]{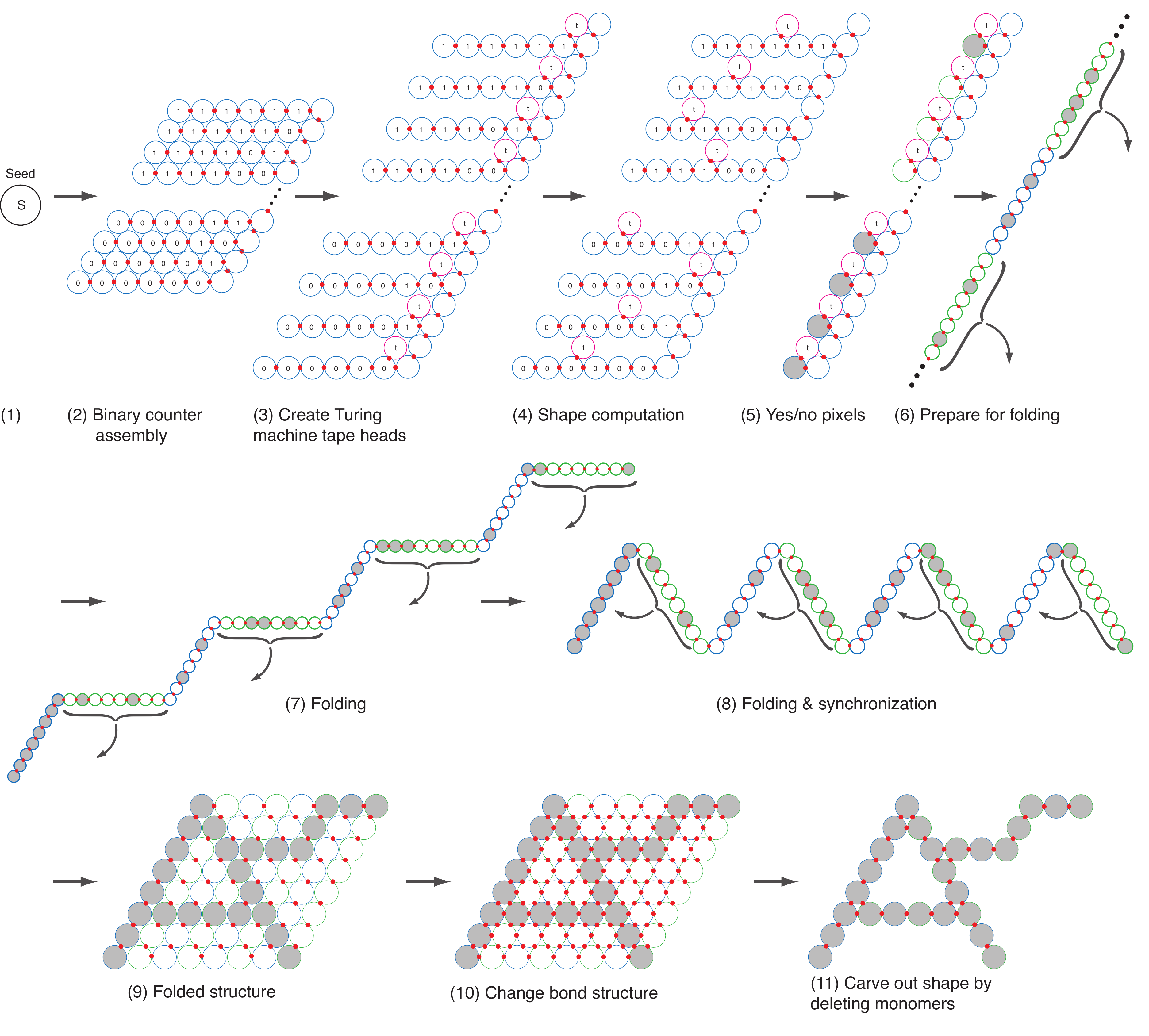}
      \caption{Construction of an arbitrary connected computable shape.}
      \label{fig:shape}
    \end{center}
  \end{figure}

\subsection{Construction overview}
Figure~\ref{fig:shape} gives an overview of the construction. We first assemble a binary counter that writes out the $n$ binary numbers $\{0,1\ldots, n-1 \}$, and where each row of the counter represents a pixel location in the $\sqrt{n} \times \sqrt{n}$ square that contains the final shape. The counter completes in expected time $O(\log^2 n)$. The counter has an additional backbone column of monomers of length $n$. After the counter is complete, each row of the counter acts as a finite Turing machine tape: the binary string on the tape represents an input $i \in \{0, \ldots n-1 \}$ to the Turing machine. For each such $i$, there is a monomer that encodes the Turing program and acts as a tape head. If the head needs to increase the length of the tape, new monomers are created beyond the end of the counter row as needed. Eventually the simulated Turing machine finishes its computation on input $i$, and the head transmits the yes/no answer to a single backbone monomer. All Turing machines complete their computation  in expected time $O(t(|n|))$, where $t(|n|)$ is the worst case time for a single Turing machine to finish on an input of length $|n| = O(\log n)$. The Turing machine head monomers then cause the deletion of the counter rows.  A synchronization on the backbone occurs after all backbone monomers encode either yes or no. The entire backbone then ``folds'' into a square, using a number of parallel ``arm rotation'' movements. Folding runs in expected time $O(\log^2 n)$. After folding, the ``no'' pixels (monomers) are deleted from the shape in a process called ``carving'', which happens in $O(1)$ expected time. After carving is complete we are left with the desired connected shape, all in expected time $O(\log^2 n + t(|n|))$ and using $O(s + \log n)$ states.

\subsection{Binary counter}\label{sec:counter}
Figure~\ref{fig:counter} gives an overview of a binary counter that efficiently writes out the binary strings that represent $0$ to $n-1\in\mathbb{N}$ in %
$O(\log^2 n)$ time,  and $O(\log n)$ states.  The counter construction builds upon the line construction in  Theorem~\ref{thm:line} (and Figure~\ref{fig:line}). The essential idea is to build a  line in one direction while simultaneously building counter rows in an orthogonal direction.  The counter begins with a single seed monomer as shown in Figure~\ref{fig:counter}(0) and ends with a configuration of the form shown in Figure~\ref{fig:counter}(8), growing  in an unsynchronized manner.

Figure~\ref{fig:counter}(1)--(6)  illustrates the construction by making the (unlikely, but valid) assumption that at configuration (1) we have created a counter that has already counted the set $\{0,1,2,3\}$, and then gives a number configurations along a trajectory to compute the set $\{0,1,\ldots,8\}$. (Note that the system is asynchronous so very few trajectories are of this nice form).  A  line is efficiently grown using the technique in Theorem~\ref{thm:line}, but where the $0$ monomers from the line are denoted with a monomer with no state name in the counter (to simplify the presentation). Starting from configuration (1), insertion events independently take place across the entire line. Each counter row in (1) is separated by unit distance, which enables multiple insertion routines to act independently (each uses a pair of monomers to form ``bridge'' monomers $b.\ast$, similar to Figure~\ref{fig:line}). Insertion of a pair of monomers triggers copying of a counter row, as  seen in configuration (4).  While a row is being copied, a 0 monomer is appended to one copy and a 1 monomer is appended to another copy.  Insertion of two monomers and the growth of the new counter row takes $O(\log n)$ expected time: this comes from the fact that insertion works in constant time, and that copying of the $O(\log n)$ monomers takes $O(\log n)$ expected time. After copying and generation of the new 0 or 1 monomers  is complete, a signal is sent to the line.  The line can then continue the insertion process. Growing a line takes expected time $O(\log n)$, but we've replaced each $O(1)$ time insertion event with an  expected time $O(\log n)$ copying process, hence the overall expected time is  $O(\log^2 n)$.

From Theorem~\ref{thm:line}, the number of states to build the backbone line is $O(\log n)$, a further $O(1)$ states can be used to carry out counter row growth and copying.

Correctness for the counter essentially follows from that of the line: We can consider that each insertion on the line is paused while counter row copying completes. After the copying the line monomers can continue their insertions, and eventually will complete. The copying and bit-flipping mechanism guarantees  that the rows of the counter encode the correct bit sequences.

\subsubsection{Counter synchronization}
\begin{figure}[t] %
    \begin{center}
      \includegraphics[width = \linewidth]{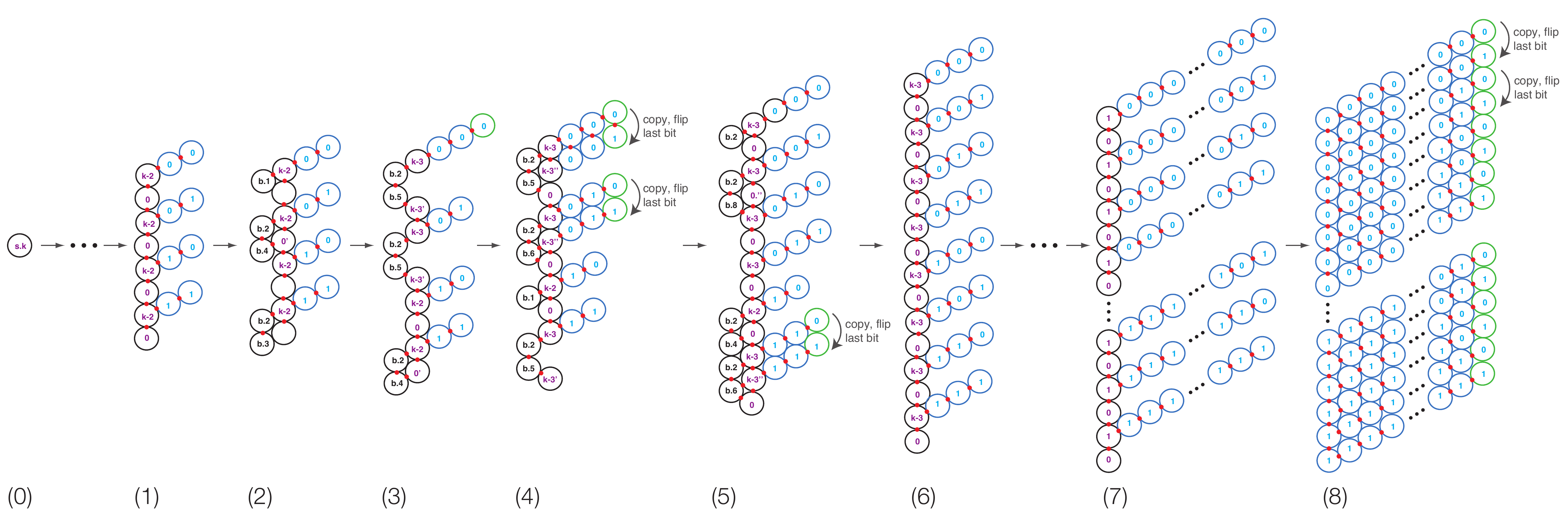}
      \caption{A binary counter that efficiently writes out the binary strings that represent all integers from $0$ to $n-1$, in $O(\log^2 n)$ time %
      and $O(\log n)$ states. The counter grows in an unsynchronized fashion, some example configurations are shown in (0)--(8), although many other trajectories are possible. A backbone line is efficiently grown using the technique in Theorem~\ref{thm:line}.  The configuration shown in (1) represents a count of 0 to 3.  In this example the backbone then expands (6) by a factor of~2, with each expansion a row of the counter is copied, with a 0 appended to one copy and a 1 appended to another copy (4). To save space the grid is rotated anti-clockwise.}
      \label{fig:counter}
    \end{center}
  \end{figure}

After the counter is complete, the backbone synchronizes.  More precisely, after the backbone has grown to its full length $n$, and all counter rows have finished their final copying (i.e. configuration (8) in Figure~\ref{fig:counter} is reached), the synchronization routine from Theorem~\ref{thm:line} is executed to inform all backbone monomers that the counter is complete,  in expected time $O(\log n)$.

\subsection{Turing machine computations}\label{sec:shape-TMs}
The next part of the construction involves simulating $n^2$ Turing machines (in parallel) in order to determine which of the $n^2$ pixels in the $n \times n$ canvas are in the desired shape, and which are not.

We begin with the synchronized counter described in the previous section. After synchronization the counter changes its bond structure so that it is of that shown in Figure~\ref{fig:shape}(2). Each row can do this independently,  by a straightforward application of Lemma~\ref{lem:chernoff}, this takes $O(\log n)$ time.

The counter then expands by a factor of 2 (in the $\vec{y}$ direction), and each row grows a single ``head'' monomer on top as shown in Figure~\ref{fig:shape}(3). The ``head'' monomer acts as a Turing machine tape head, as in Figure~\ref{fig:TM-example}, and treats the counter row as a finite Turing machine tape. The input to the Turing machine is the binary number $i \in \{ 0,1,\ldots, n^2-1\}$ stored on the tape.  The Turing machine program executed by the head is of size $s$ and is stored explicitly in the rules that are applicable to the head monomer, this gives the~$O(s)$ term of monomer states in the statement of Theorem~\ref{thm:shape}.  If the Turing machine requires more than $\log_2 n$ tape space  (the length of the counter row),  then new monomers are grown to the left as required (we assume that the turing machine runs on a single, one-way infinite tape). When the Turing machine enters an accept or reject state, the head moves to the right, deleting each tape monomer along the way, and communicates this bit to the backbone row, as shown in Figure~\ref{fig:shape}(5). Each tape head then deletes itself, and $n^2$  backbone monomers undergo a synchronization when all tape head monomers are deleted. We are left with $n^2$ monomers as shown in Figure~\ref{fig:shape}(6), each of which stores a bit representing whether or not it represents a pixel in the final desired shape. 

We have~$n^2$ Turing machines to simulate in parallel. In the folding step below we modify the Turing machine program so each machine first computes a simple inequality, which causes each simulated Turing machine to run for time $t(n) = \Omega(\log_2 n)$.   Then, by setting $m = n^2$, and $k = t(|n|)$ in Lemma~\ref{lem:chernoff}, all Turing machines finish their computations in expected time $O(t(|n|))$.

\subsection{Folding}
In this part of the construction the line of $n^2$  monomers folds itself into an $n \times n$ square. The folding process is outlined in 
Figure~\ref{fig:shape}(7)--(9), and is accomplished as follows. 

In the previous section, we used a Turing machine computation on the $i^{\mathrm{th}}$ row of a counter was to decide whether or not  pixel $i \in \{ 0,1,\ldots n^2\}$ is in the final shape. We use these same Turing machines to carry out an additional computation. We can see in Figure~\ref{fig:shape}(6)--(9) that the line of monomers is divided into alternating segments, each of length $n$, that fold in one of two ways. In particular, the monomers highlighted in green each carry out a sequence of 3 clockwise rotations with respect to their left neighbor. This can be done using rules similar to the single rule in the arm rotation example shown in Figure~\ref{fig:rotation}.  For each monomer to know whether it should rotate (green) or not (blue), we have the Turing machine check if $i$ satisfies the inequality  $ (2j-1)n  \leq i < 2jn $, and if so this bit is communicated to the $i^{\mathrm{th}}$ backbone monomer (along with the yes/no pixel information). Then after the synchronization described in the previous section, monomers with indices $i$ that satisfy the inequality rotate as shown  in Figure~\ref{fig:shape}(6)--(9). A synchronization takes place for each rotating arm, after the second rotation of each monomer, as shown in Figure~\ref{fig:shape}(8).

The expected time for the Turing machine monomer head to compute the inequality is $\Omega(\log n)$, as it requires reading the entire input (given our orientation of binary strings), and this is already accounted for in Section~\ref{sec:shape-TMs} above. The expected time to rotate an arm (twice) so that it is in the position shown in Figure~\ref{fig:shape}(8) is $O(\log n)$, for each individual arm. By letting $m=n^2 / 2$ and $k=\log n$ in Lemma~\ref{lem:chernoff}, the expected time for all $n$ arms to rotate to the configuration shown in Figure~\ref{fig:shape}(8) is $O{(\log n)}$. Then we apply $n$ synchronization steps, each of which runs in expected time $O(\log n)$, giving an expected time of  $O(\log^2 n)$. As with the first rotations, the final rotations occur in expected time $O(\log n)$, to give a total expected time of $O(\log^2 n)$ for the folding step. The number of states used for folding is $O(1)$.

\subsection{Bonding and carving}
The goal here is to delete those pixels that should not be in the final stage, and to do this in a way such that the shape does not become unintentionally disconnected (otherwise if we delete too early, e.g. before all ``arms'' have folded or all bonds have formed, part of the shape could become disconnected). In Figure~\ref{fig:shape} ``yes'' pixels belonging to the final shape are highlighted in gray, and ``no'' pixels that should be deleted are in white. After folding, i.e.\ when green and blue outlined monomers begin to come into horizontal ($\vec{x}$) contact, as shown in Figure~\ref{fig:shape}(9), these monomers bond to all of their neighbors. Then the following distributed {\em carving} algorithm is executed, a ``no'' pixel is deleted if and only if either (a)  it is bonded to its 6 neighbors and they have the maximum number of bonds to each other  (b) all neighbors that have not been deleted already are bonded to it and to each other. 

This procedure prevents the shape from becoming disconnected by the following argument. Any two adjacent regions, one containing as yet unbonded monomers, and the other containing deleted monomers are bordered by a connected path of rigid bonds. The only way for monomers on this rigid path to be deleted are if neighboring unbonded monomers become bonded, thereby maintaining the connectedness of the boundary and making more monomers available for safe deletion.

Assuming folding has completed, bonding is a local operation that for each individual monomer runs in time $O(1)$. The same is true for carving. The expected time for all monomers to bonding and carving is $O(\log n)$.

This completes the construction for Theorem~\ref{thm:shape}.

\section{Efficient computation of patterns}\label{sect:pattern} 

Let $|n| = \lceil \log_2 n \rceil$.
\begin{theorem}\label{thm:pattern}
An arbitrary finite computable 2D pattern of size $\leq n \times n$, where $n = 2^{p}, p\in \mathbb{N}$, with pixels whose color is computable on a  polynomial  time $O(\log^{\ell} n)$ (inputs are of length $O(\log n)$), linear space  $O(|n|)$, program-size $s$ Turing machine, can be constructed  in expected time %
$O(\log^{\ell+1} n)$, with $O(s + \log n)$ monomer states and without growing outside the pattern borders. Moreover,  this can be done  without explicitly using synchronization.
\end{theorem}

  \begin{figure}[t]
    \begin{center}
      \includegraphics[width = \linewidth]{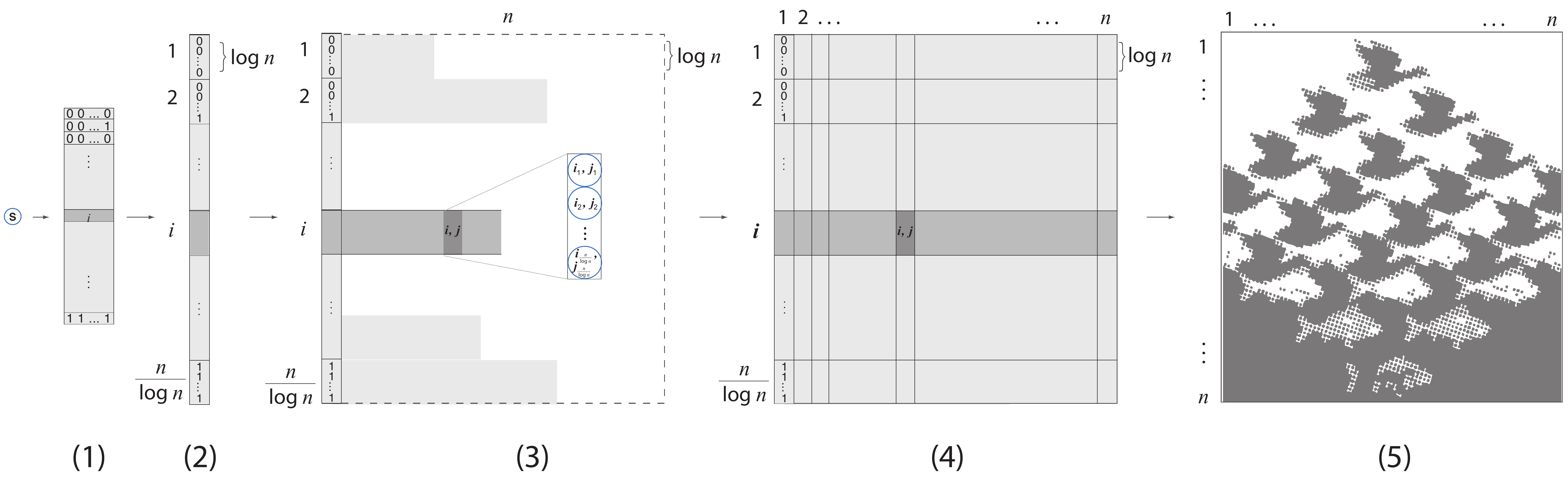}
      \caption{Construction of an $n \times n$ pattern in expected time polylogarithmic in $n$, without growing outside the pattern borders. No long range communication (synchronization) is used and so the computation happens in a completely asynchronous and distributed fashion. Final configuration adapted from M.\ C.\ Escher, Sky \& Water I, woodcut, 1938.}
      \label{fig:patternoverview}
    \end{center}
  \end{figure}

The construction is described in the remainder of this section.  Figure~\ref{fig:patternoverview} gives an overview. For simplicity, the figure is drawn on a square grid.  It should be noted that the construction occurs in a completely asynchronous distributed fashion. For ease of exposition, we first describe a construction where $n$ is a double power of 2, i.e.\ $n = 2^{2^p}$ for $p\in \mathbb{N}$, and then in  Section~\ref{sec:patternPower2} we describe how to modify it for $n = 2^p$.

  \begin{figure}[h]
    \begin{center}
      \includegraphics[width = 0.4\linewidth]{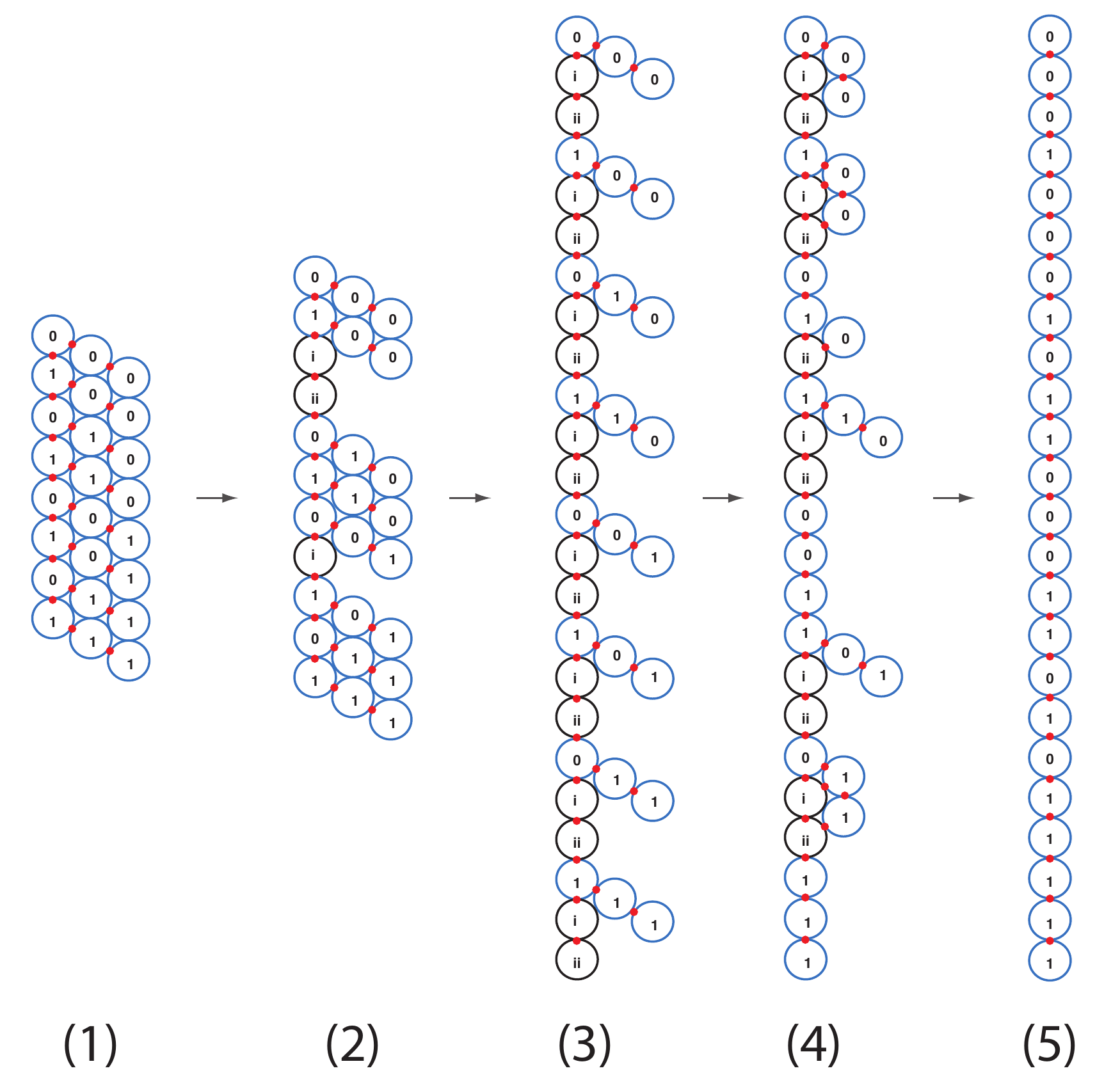}
      \caption{Sketch of a procedure to efficiently reconfigure a counter into a straight line. }
      \label{fig:patternfolding}
    \end{center}
  \end{figure}

  \begin{figure}[h]
    \begin{center}
      \includegraphics[width = 3 in]{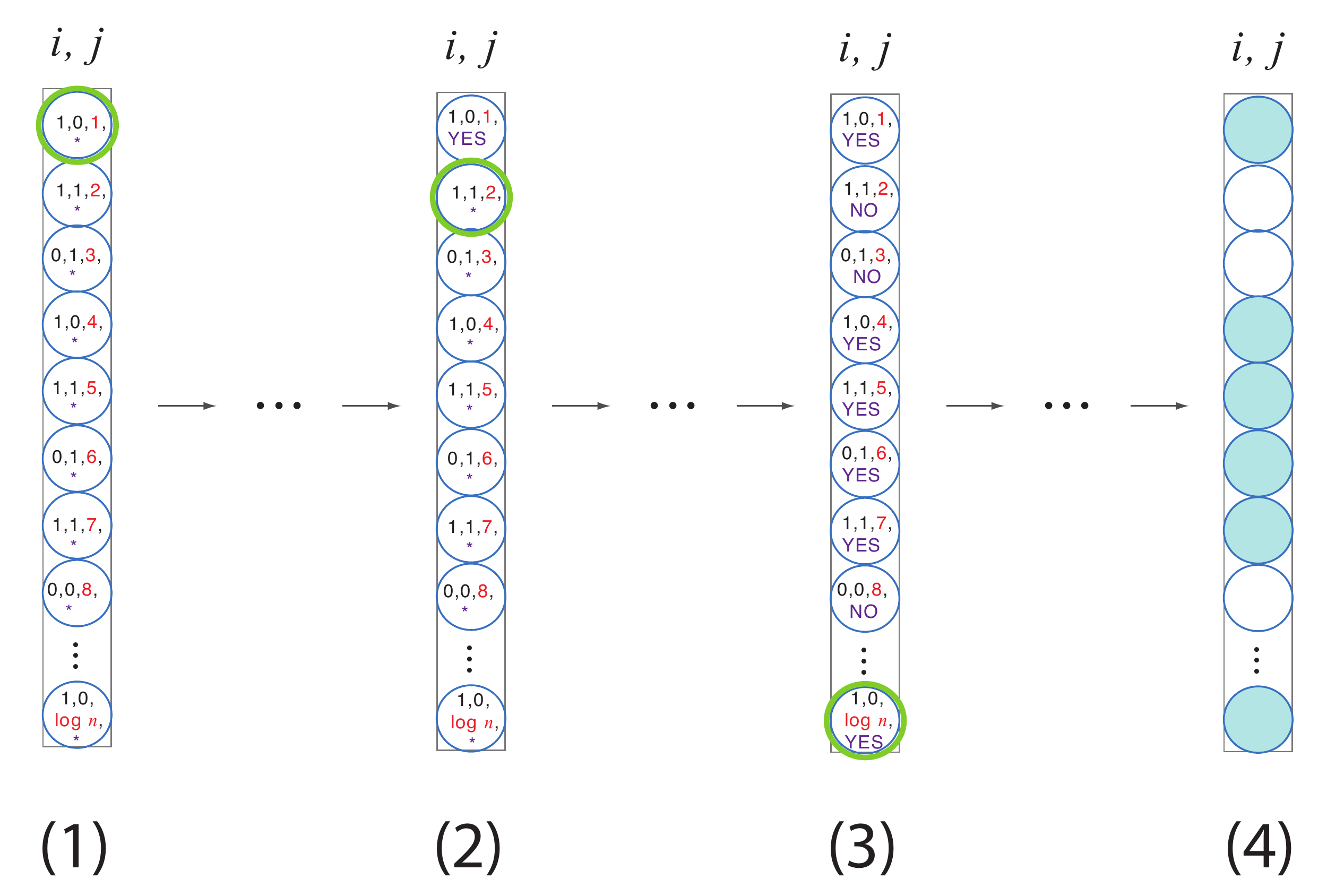}
      \caption{A strip of monomers of length $\log_2 n$. (1) The entire strip acts as an input tape and work tape for a Turing machine that uses $\log_2 n$, whose head is denote in green. (2) After the first Turing machine has completed its computation the second machine is initialized, and so on until machine $\log_2 n$ completes its computation. (4) The output, yes or no, from each machine is denoted as solid green or white.}
      \label{fig:patternTM}
    \end{center}
  \end{figure}

\subsection{Vertical and horizontal binary counters}
The construction begins by growing a counter, shown in Figure~\ref{fig:patternoverview}(1), that contains all binary numbers from $1$ to $n / \log_2 n$. The counter is  described in Section~\ref{sec:counter}, and illustrated in Figure~\ref{fig:counter}. The only difference here is that we omit the final synchronization steps from Section~\ref{sec:counter}. Notice that such a counter contains exactly $n$ monomers; i.e.\  $n / \log_2 n$ rows each of length  $\log_2 (n /\log_2 n)$. The counter rows then expand so that they are of length $\log_2 n$ (we omit the details, but this is easy to achieve with the number of states permitted in the theorem statement). As counter rows finish,  we want to rotate them so that they stand in a column as shown in Figure~\ref{fig:patternoverview}(2), although due to the asynchronous nature of the computation they will most likely not do this as shown.  As can be seen in Figure~\ref{fig:counter}, the counter is stable throughout its entire construction.
 
Figure~\ref{fig:patternfolding} shows how this is accomplished. After a counter row completes, the backbone monomer that is attached to that counter row expands vertically by $\log_2 n$ monomers. Then the counter row rotates from a horizontal, to a vertical position as shown in  Figure~\ref{fig:patternfolding}. Finally the binary string stored in the counter row is copied to the expanded backbone monomers, and the counter row deletes itself. All of this is carried out while maintaining stability (Definition~\ref{def:stable}). By the time all rows rotate we have a backbone of height $n$. 

After row $i$ rotates to the vertical orientation, it immediately initiates growth of a second, horizontal, counter which counts while copying the binary number $i$. Note that~$i$ can not be stored in single monomer (as this would require $\Omega(n)$ states for the entire construction), instead the entire {\em strip} of $\log_2 n$ monomers encoding $i$ is copied as the second counter grows.  A sketch is shown in row $i$ of Figure~\ref{fig:patternoverview}(3). This second counter works as follows. Its backbone line grows horizontally, starting from the top monomer of the $i$ strip. During each insertion event, the $\log_2 n$ monomers that encode $i$ are copied. At the same time  these $\log_2 n$ monomers are used to encode the values being generated by  second counter (essentially, while copying $i$ we execute the copying and bit flip idea seen in Figure~\ref{fig:counter} to get a new value $j$). By the time row $i$ finishes it contains $n$ vertical strips of $\log_2 n$ monomers, and each strip  encodes a distinct pair $(i,j)$ where $ j \in \{ 0,1,\ldots n -1\}$.

To find the expected time to complete all strips notice that each strip $(i,j)$ can grow independently from all others, and each strip depends on only $O(\log^2 n)$ events to happen in order for that strip to be complete. %
Hence we can apply Lemma~\ref{lem:chernoff} by setting $m = n^2 / \log n$ (the number of strips), and $k = O(\log^2 n)$, to get an expected time of $O(\log^2 n)$ for all strips to complete.

\subsection{Turing machine computations}
We will treat each strip $(i,j)$ as a Turing machine tape. After a strip completes, a signal is sent from the top to the bottom of the strip, successively writing out one of the integers $p \in \{1,2,\ldots, \log_2 n \}$ in each of the $\log_2 n$ monomers. This signal also tells the topmost monomer  that it now encodes a Turing machine tape head, as shown in green in Figure~\ref{fig:patternTM}(1).  The encoded tape head moves up and down the strip as required. By the theorem hypothesis, the Turing machine requires at most $\log_2 n$ space, which is exactly the strip length. These Turing machine are assumed to takes as input three integers $(i,j,p)$ which provide a unique coordinate for each monomer in the entire $n \times n$ pattern. In time polynomial in its input length $|n| = \log_2 n$, the first Turing machine decides whether pixel $(i,j,1)$ is black or white, communicates this bit to the topmost monomer, and moves on to the second from top monomer, which has coordinate $(i,j,2)$. This process continues until all $\log_2 n$ monomers in the strip are colored either black or white.

Each  Turing machine runs in time  $O(\log^\ell n)$, which takes expected time $O(\log^\ell n)$ to simulate. We can apply the Chernoff bound in Lemma~\ref{lem:chernoff} by setting $m= n^2 / \log n$ (the number of strips) and $k=  O( \log^{\ell+1} n)$ (the expected time for all $\log_2 n$ Turing machines to finish on a single strip) to get an expected running time of  $O( \log^{\ell+1} n)$ for all Turing machine computations to complete.

The overall time bound for the entire construction is $O(\log^2 n + \log^{\ell+1} n) = O(\log^{\ell+1} n)$ since we know $\ell \geq 1$ (because our Turing machines are required to read their entire input). %

\subsection{Patterns with diameter a power of 2}\label{sec:patternPower2}
The previous construction works for when $n$ is a double power of 2. The following text shows how to modify the construction so it works for $n$ being a single power of 2,  i.e.\ $n = 2^p \in \mathbb{N}$. In this case, the counter that builds Figure~\ref{fig:patternoverview}(1) is modified so that it does not produce the final (bottom) row with index $\lceil n / \log n \rceil$. Let $k = n - \lfloor n / \log n \rfloor \log n$, and observe that $k < \log n$. Now, the bottoms row has  index $b = \lfloor n / \log n \rfloor$. Row $b$ triggers growth of a counter as before, however when each column (of length $\log_2 n$) of the counter finishes, it grows an additional $k$ monomers, each with a unique id $i \in \{1,\ldots,k\}$. This results in a counter of size $n \times (k + \log n)$. When all counters complete in the entire construction, the canvas is of size $n \times (k + \lfloor n / \log n \rfloor \log n) = n \times n$. The Turing machine computations proceed as follows. As each column of the counter in row $b$ completes, the $\log n$ Turing machine computations in that column proceed as before. When they are finished, an additional $k$ Turing machine computations are triggered, which run one after the other, and use  $k + \log n$ workspace in the column (which is more than enough), and the unique IDs of the $k$ extra monomers in order to decide whether or not each of those $k$ pixels belong in the pattern.  These additional aspects of the construction merely add a constant factor to the time analysis.

\section{Discussion and future work}\label{sect:discussion}
We have introduced a model of computation called the nubot model, and explored its ability to construct shapes and patterns. We have shown that the model is capable of efficiently generating a wide variety of shapes and patterns exponentially quickly. The intention for our model is to explore the abilities of molecules to compute in ways that are seen in nature and that we are starting to see in the laboratory.  This perspective leaves a lot of room for future work.

One interesting direction  is to explore the algorithmic limits of {\em dynamic} structures. In this paper we have used our active self-assembly model to grow shapes and patterns that are ultimately static. One can also consider shapes and patterns that are forever dynamic. Cellular automata are a well-studied model from this point of view, although they are incapable of expressing our notion of movement and active self-assembly. What kinds of dynamic structural  systems can, and can not, be modeled by nubots?  

One possible objection to our model, on physical grounds, is the lack of any realistic notion of persistence length (which is also absent from many models of self-assembly). On the one hand, it is clear that there are natural and artificial structures of high aspect ratio coupled with high tensile strength or persistence length (hair, actin filaments, microtubules). On the other hand, `high' does not mean `infinite'! One could introduce complexity measures of tensile strength or persistence length and analyze the capabilities of the nubots model with respect to these resources. The important point here would be to appropriately define these measures so that they capture what is observed at the molecular scale in the laboratory. 

For the topic of tensile strength, 
one could take inspiration from 
cellular migration and adhesion in developmental biology: nubot monomers could have variable strength bonds which break if there is enough movement in one direction. For example, bonds could have strength $s \in [0,1] \subseteq \mathbb{R}$ where 0 is not bonded, 1 is fully bonded, and other values are of intermediate strength.  Objects are pulled apart if enough movement rules are applicable and so that their bond strength, or tensile strength, can not overcome the strength of movement. 
What are the classes of systems that can and can not be built under such constraints? If we have to pay for bond strength, in general, is it possible to place bounds on this cost in terms of the shapes, patterns or dynamics we wish to model?

As noted in the introduction, the field of reconfigurable robotics considers a wide range of models that share a number of features with our nubot self-assembly model. In particular, reconfigurable modular robots with a similar long-range movement primitive to ours can achieve arbitrary reconfiguration in time logarithmic in shape size~\cite{Crystalline_ISAAC2008}, and are capable of linear parallel time reconfiguration with more realistic physical constraints~\cite{aloupis2011efficient}. It remains as future work to compare such models to ours. What would be particularly interesting would be to explore the differences in model capabilities that are solely due to the inherent differences in macro-scale and molecular-scale self-assembly (in molecular systems we are typically unconcerned with gravity and friction; energy in the form of fuel molecules may be freely available in the environment; temperature, brownian motion and other forms of agitation play a major role). 

There are also some more technical questions arising from our work. We use a Chernoff bound in Lemma~\ref{lem:chernoff} to simplify the time analysis of our constructions. However, many assembly systems that are expressible in our model do not satisfy the conditions of this lemma. It would be nice to find other tools, perhaps more general, to aid in the time analysis of nubots systems. For example, starting from a single monomer, if the longest sequence of rule applications that can lead to some `terminal' monomer type is $k$, then is $\tilde{O}(k)$ the expected time for all rules to complete? 

The pattern construction in Section~\ref{sect:pattern} terminates with an assembly that is not completely connected (however, it is stable and connected). In that construction we intentionally did not use synchronization over long distances, but by using synchronization it is indeed possible (and relatively easy) to modify the final pattern so that it is completely connected. However, given that synchronization has such power, it is interesting to ask what can be done in its absence.  Without using synchronization, or any similar form of rapid communication over long ($> \log n$) distances, is it possible to deterministically assemble an $n\times n$ completely-connected square in time polylogarithmic in $n$?  

Without the movement rule, the nubot model is a kind of asynchronous and nondeterministic cellular automaton. Thus in the absence of movement in our model, cellular automata are a good starting point to assess its computational complexity  (how efficiently can problems be solved?). However, with movement, it is clear that our model can carry out certain tasks that cellular automata, or indeed Turing machines, can not (even under reasonable encodings). What are the upper bounds and  lower bounds on the computational complexity of our model? For example,  Section~\ref{sect:simulation} gives a polynomial (in nubots time plus number of monomers) time algorithm for simulating a nubots trajectory. If nubots time is merely polylogarithmic in the maximum number of monomer types, then is it possible to simulate nubots in polylogarithmic time on a parallel computer (e.g. on polylogarithmic-depth Boolean circuits)?

One could consider variations on the rules. Already it is the case that movement rules facilitate non-local interactions. One could take this even further, as we now discuss. Consider a variant on the movement rule  where the application of a movement rule $r$ remotely triggers the application of another movement rule $r'$. More precisely, let $A$ be an arm monomer bonded to base monomer $B$, and let  $\mathcal{M}(\config, A, B, \vec{v}) \neq \{\}$ be the  movable set for the application of rule $r$ to monomers $A,B$. Also, in the same configuration $\config$, there is another pair of bonded monomers $D,E$, with an applicable movement rule $r'$, and where $D \in \mathcal{M}(\config, A, B, \vec{v}) $ and $E \notin \mathcal{M}(\config, A, B, \vec{v}) $, and where $r'$ translates $D$ by $\vec{v}$. When $r$ is applied it triggers the application of $r'$: i.e.\ when monomer $D$ moves because of the remote application of $r$, both $D$ and $E$ change state as if rule $r'$ was applied; hence $r$ and $r'$ are applied at the same time.  We do not consider this kind of remote rule triggering  in this paper, but we mention it here as a way for potential future work to model non-local interactions that can occur due to movement. %
This is one possible way to model systems where interactions  occur in a decentralized  manner. %

The model uses both rigid and flexible bonds. Besides simple examples given early in the paper, the only construction that uses flexible bonds is synchronization, and it turns out that one can design a synchronization routine that does not use flexible bonds that works in the presence of agitation (by building a rigid row immediately below, and parallel to, the synchronization row---see Figure~\ref{fig:sync}---that stops monomers floating away). What classes of (perhaps scale-invariant) shapes can be assembled by exploiting flexible bonds that can not be assembled otherwise?  

Finally, since the model is directly inspired by a wide election of natural and artificial molecular  systems, this begs the question: can nubots be implemented at the molecular scale in the laboratory?

\section*{Acknowledgments}
Many thanks to Niles Pierce and Patrick Mullen for valuable discussions and input.  We also thank Moya Chen, Doris Xin, Joseph Schaefer, and Andrew Winslow for stimulating and fruitful discussions. 

%
%

% \dw{\input{armsfingerscounter}}  % old counter not included
\section*{References}
%\bibliographystyle{abbrv} % plain
%\bibliography{nubots}

\end{document}